\documentclass[a4paper,11pt]{article}

\usepackage{jcappub}
\setcitestyle{numbers,sort&compress}
\usepackage{caption}
\usepackage{wrapfig}
\usepackage{aas_macros}
\usepackage[skip=0.5ex]{subcaption}
\usepackage{mathbbol}
\usepackage{slashed}
\usepackage{xcolor}
\usepackage{enumitem}
\usepackage{cleveref}
\usepackage{tikz}
\usepackage{orcidlink}
\usetikzlibrary{arrows.meta, positioning, calc}
\usepackage{amsmath, amssymb}
\usepackage{lineno}

\definecolor{kszred}{RGB}{180, 60, 50}
\definecolor{lensblue}{RGB}{50, 100, 180}
\definecolor{velteal}{RGB}{30, 140, 120}
\definecolor{tmplpurple}{RGB}{120, 70, 160}
\definecolor{resultgreen}{RGB}{50, 130, 60}
\definecolor{labelgray}{RGB}{120, 120, 120}

\title{
Direct shear $\times$ kSZ correlation: 
controlling baryons without modeling galaxies
}

\author[a]{Anoma Ganguly,\orcidlink{0000-0002-5915-4245}}
\author[b, c]{Emmanuel~Schaan,\,\orcidlink{0000-0002-4619-8927}}
\author[a,d]{Elisabeth Krause, 
\orcidlink{0000-0001-8356-2014}}
\author[a,d]{Tim Eifler
\orcidlink{0000-0002-1894-3301}}

\affiliation[a]{Department of Astronomy, University of Arizona, Tucson, AZ 85721, USA}
\affiliation[b]{SLAC National Accelerator Laboratory, 2575 Sand Hill Road, Menlo Park, California 94025, USA}
\affiliation[c]{Kavli Institute for Particle Astrophysics and Cosmology, Stanford University, 452 Lomita Mall, Stanford, CA, 94305, USA}
\affiliation[d]{Department of Physics, University of Arizona, Tucson, AZ 85721, USA}
\emailAdd{anomaganguly@arizona.edu}

\abstract{
Baryonic feedback redistributes gas within and around dark matter haloes, suppressing the small-scale matter power spectrum at a level that is now the leading systematic for upcoming weak-lensing surveys. 
The kinetic Sunyaev--Zeldovich (kSZ) effect directly probes this redistributed gas, but existing measurements around galaxies are either tied to the properties of the chosen galaxy sample or are susceptible to biases from other extragalactic foregrounds. 
We address both by cross-correlating a kSZ template constructed from the tomographic weak-lensing convergence maps and the radial velocity maps reconstructed from galaxy surveys via the continuity equation, with the observed CMB temperature.  
We forecast the detectability for Rubin LSST~Y10 and Roman kinematic lensing samples combined with ACT, SO, and CMB-HD, finding signal-to-noise ratios of $\sim 5$–$15$ for current and upcoming CMB data and $\gtrsim 100$ for CMB-HD, corresponding to few-percent and sub-percent constraints, respectively, on the baryonic suppression of the matter power spectrum. This method should thus achieve sufficient statistical precision to model baryonic feedback effects for Rubin, without the systematic challenge of modeling any galaxy--matter connection.
}

\begin{document}
	{\tiny {\tiny}}	\maketitle
	\flushbottom
\section{Introduction}
The next generation of weak lensing surveys such as the Rubin Observatory Legacy Survey of Space and Time (LSST) \cite{Ivezic2019}, Euclid \cite{Laureijs2011}, and Roman (Roman Space Telescope) \cite{Spergel2015} is expected to reach percent-level statistical precision on cosmic shear across non-linear scales. Extracting unbiased cosmological constraints from these measurements therefore requires the non-linear matter power spectrum to be predicted at comparable, percent-level accuracy. On these scales, the astrophysical processes from active galactic nuclei and supernovae redistribute baryons within and beyond their host dark matter haloes, resulting in a suppression in the matter power spectrum by up to tens of percent at $k \simeq 1\,h^{-1}\,$Mpc compared to the dark-matter-only predictions. Current frameworks for modeling baryons rely on either semi-analytic prescriptions \cite{SchneiderTeyssier2015, Mead2020, Arico2021} or hydrodynamical simulations with sub-grid physics \cite{vanDaalen2011, Vogelsberger2020, Navarro2021,
Schaye2023}, whose predictions disagree at the tens-of-percent level on small scales ($k\gtrsim1\,h\,\mathrm{Mpc}^{-1}$) \cite{Chisari2019, Huang2019, vanDaalen2020, Crain2023}.
This in turn motivates new observables that are robust and can constrain the  baryon distribution directly from observations.

\begin{figure}[t]
    \centering
    \includegraphics[width=1.0\linewidth]{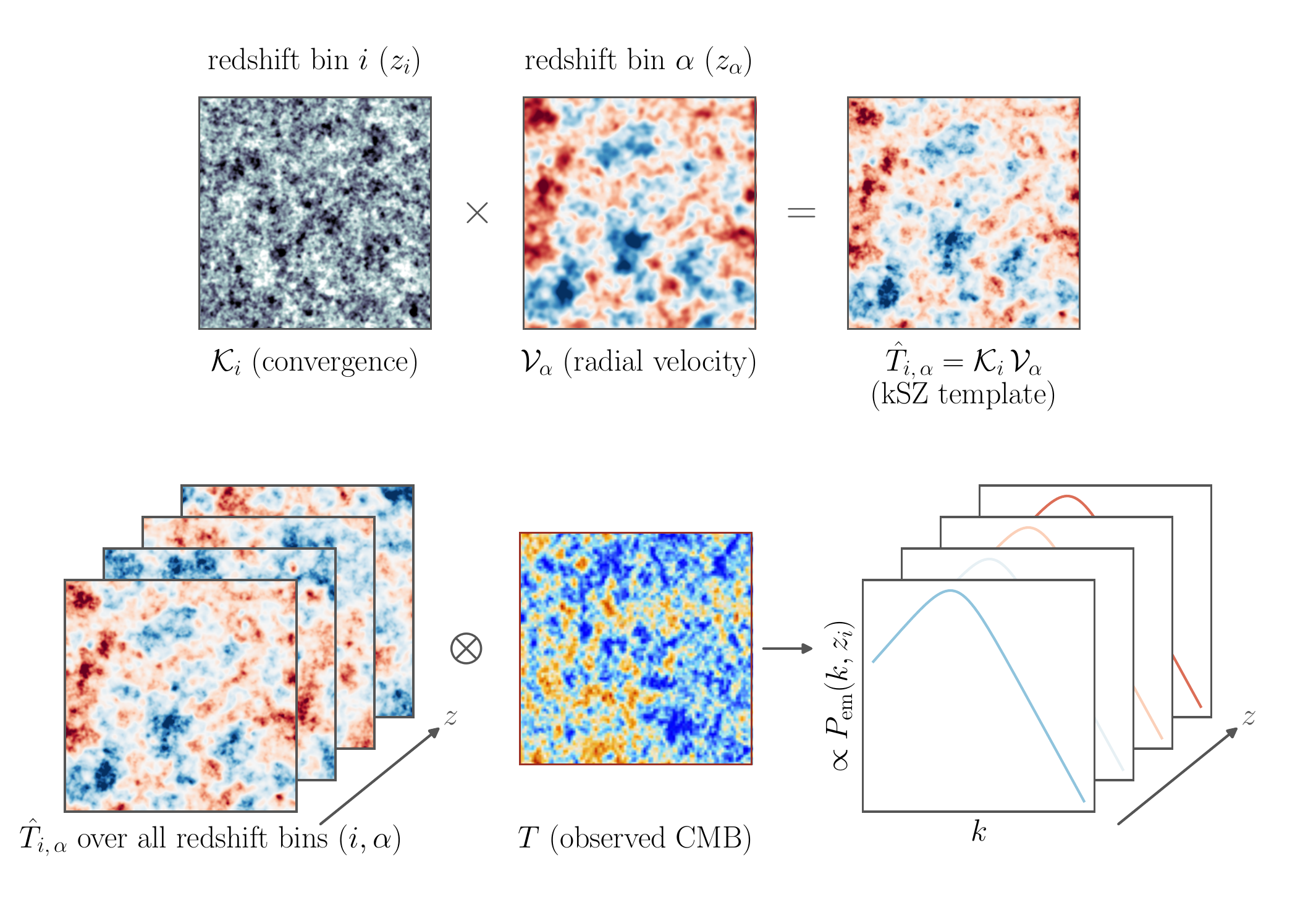}
    \caption{A schematic diagram illustrating how the cross-correlation of the kSZ signal (Sec.~\ref{subsec:ksz}) with weak lensing (Sec.~\ref{subsec:weaklensing}) and reconstructed velocity field (Sec.~\ref{subsec:recons_vel}) probes the baryonic contribution to the matter power spectrum. \emph{Top row:} The weak-lensing convergence $\mathcal{K}_i$ (tracing the projected matter density) in redshift bin $z_i$ is multiplied by the projected radial velocity field $\mathcal{V}_\alpha$ in redshift bin $z_\alpha$, forming the kSZ template $\hat{T}_{i,\alpha} = \mathcal{K}_i\,\mathcal{V}_\alpha$. \emph{Bottom row:} The kSZ templates built across all redshift-bin combinations $(i,\alpha)$ are cross-correlated ($\otimes$) with the observed CMB temperature map $T$. These templates when cross-correlated with $T$ extract the kSZ contribution and are sensitive to the electron–matter cross power spectrum $P_{\mathrm{em}}$.}
    \label{fig:visualization}
\end{figure}
\begin{figure}[t]
    \centering
    \includegraphics[width=0.6\linewidth]{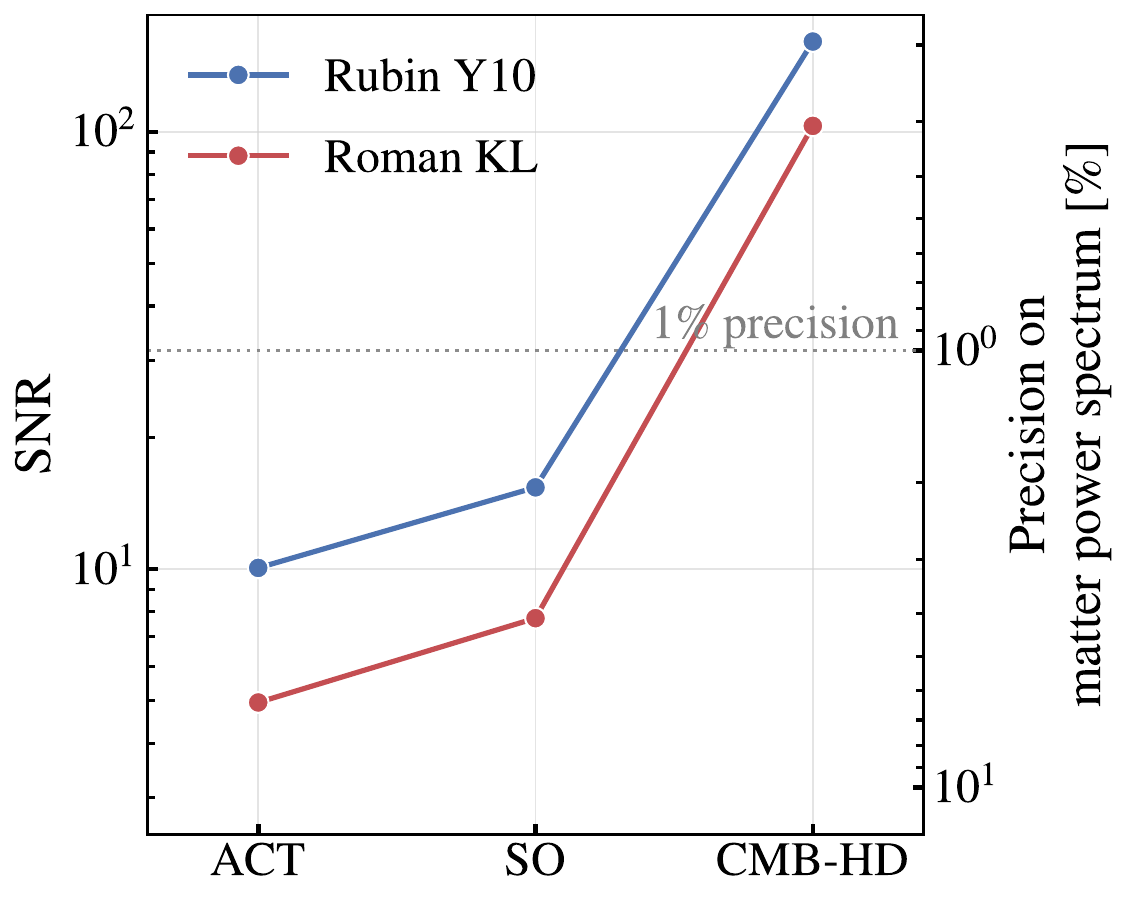}
    \caption{Forecast SNR for the kSZ template cross-correlation, for two weak-lensing surveys (Rubin LSST Y10, Roman KL) each combined with three CMB experiments (ACT, SO, CMB-HD). The right axis converts SNR into an approximate precision on the matter power spectrum ($\approx 2(\Omega_b/\Omega_m)\,/\mathrm{SNR}$ using Eq.~\eqref{eq:Pmm-split-fractional}). Both surveys reach a few percent precision for ACT and SO and sub-percent for CMB-HD, comparable to the accuracy needed to model baryonic feedback for next-generation weak-lensing surveys.
    }
    \label{fig:max_snr}
\end{figure}

The kSZ effect is one such probe that traces the electron distribution directly is the kSZ effect. It arises when the Cosmic Microwave Background (CMB) photons Thomson-scatter off ionized electrons with a bulk velocity along the line of sight \cite{Sunyaev1980}. Efforts to measure this signal have advanced considerably in recent years: pairwise estimators \cite{Hand2012, Soergel2016, Calafut2021},  velocity-reconstruction stacking \cite{Schaan2021, Hadzhiyska2024, Qu2026},
and projected-field methods \cite{Hill2016, Kusiak2021, Bolliet2023} 
have all yielded significant detections with signal-to-noise improving as CMB and large-scale structure datasets continue to grow.

The kSZ temperature fluctuation is proportional to the product of the electron density and the line-of-sight peculiar velocity. Since the line-of-sight velocity averages to zero, the two-point cross-correlation of a kSZ map $T$ with a matter density tracer $g$, $\langle T g\rangle$, vanishes. 
A non-zero signal therefore requires at least a three-field statistic, which can be built in two possible ways \cite{Smith2018}. 
The pairwise kSZ, kSZ stacking and velocity-reconstruction measurements belong to the class of $\langle ggT\rangle$-type estimators that reconstruct the line-of-sight velocity from the galaxy field and correlate it linearly with the CMB temperature
\citep{PhysRevLett.109.041101,PhysRevLett.115.191301,Bernardis_2017,Soergel_2016,Sugiyama_2018,PhysRevD.104.043502,li2024detectionpairwisekineticsunyaevzeldovich,Hadzhiyska:2025egz,harscouet2025kszeveryonepseudoclapproach,PhysRevD.93.082002,PhysRevD.103.063513,PhysRevD.108.023516,hadzhiyska2025evidencelargebaryonicfeedback,guachalla2025backlightingextendedgashalos, mccarthy2024atacamacosmologytelescopelargescale,mccarthy2025atacamacosmologytelescopecrosscorrelation,hotinli2025velocityreconstructionkszmeasuring,lai2025kszvelocityreconstructionact,lague2024constraintslocalprimordialnongaussianity,2026arXiv260419744Q, 2026arXiv260419745H}. 
Projected-field estimators of the $\langle T^2g\rangle$-type instead cross-correlate the squared CMB temperature with the galaxy overdensity \citep{PhysRevLett.117.051301,PhysRevD.94.123526,PhysRevD.104.043518,Bolliet_2023}.
Squaring the temperature converts foregrounds (tSZ, CIB, radio sources) and CMB lensing into biases that survive the cross-correlation, requiring aggressive component separation.

In both cases, the statistical precision is very high, but the interpretation in terms of baryon contribution to cosmic shear is limited by challenging systematics:
(1) the galaxy sample used only probes halos over a limited range of mass and redshift outside of which extrapolation is needed, and (2) modeling the host halo population of these galaxies is difficult.
Indeed, galaxies can be centrals or satellites in halos with complex mass distributions, miscentered with respect to the halo center \cite{2026arXiv260304397K, 2021ApJ...919....2M, 2023arXiv230710919M}.
Their connection to the host halo may depend on properties beyond mass or formation histories (assembly bias).
All these factors may bias the interpretation of kSZ measurements around galaxies, including comparisons with simulations \citep{PhysRevD.94.123526,hadzhiyska2025evidencelargebaryonicfeedback,Sunseri:2025hhj,Bigwood:2025kur,bigwood2025kineticsunyaevzeldovicheffect,Siegel:2025ivd,salcido2025implicationsfeedbacksolutionss8,mccarthy2025flamingocombiningkineticsz,Lague:2025txe,kovac2025baryonificationiiconstrainingfeedback}.
This is a major challenge for all existing analyses that aim to address baryon uncertainties in cosmic shear via joint analyses with kSZ measured around galaxies \cite{Schneider_2022, DES:2024iny, Hadzhiyska2024, bigwood2025kineticsunyaevzeldovicheffect, Lague:2025txe,kovac2025baryonificationiiconstrainingfeedback}.

In this work, we replace one of the galaxy fields with the weak-lensing convergence in the $\langle ggT\rangle$-type estimator. 
This completely avoids the issue of modeling the galaxy--matter connection and the extrapolation outside of the mass and redshifts probed by the galaxy samples.
By probing the matter--gas power spectrum, this method directly measures the baryon contribution to cosmic shear.
Analogous approaches have been proposed with fast radio bursts rather than kSZ \cite{2025arXiv250919514L}.
This estimator is constructed by multiplying the convergence map by a radial-velocity map reconstructed from the large-scale galaxy distribution \cite{PhysRevD.109.103533, PhysRevD.109.103534}. 
Being linear in the line-of-sight velocity, the resulting kSZ template, when cross-correlated with the observed CMB temperature, isolates the kSZ contribution. The primary CMB and foregrounds, uncorrelated with the reconstructed velocity, average away. 
Since the convergence field traces the projected total matter and galaxies enter only through velocity reconstruction on large, linear scales, the estimator is directly sensitive to the electron--matter power spectrum $P_{\rm em}$. 
Being linear in $T$, it is also far less susceptible to foregrounds than $\langle T^2g \rangle$-type estimators.

Weak-lensing kernels are broad, extending from the source galaxy all the way to the observer. 
Thus a naive implementation would average feedback over the full line of sight. 
We therefore apply the Bernardeau Nishimichi Taruya (BNT) nulling transform \cite{HutererWhite_2005, Bernardeau_2014} to construct lensing kernels localized in comoving distance and correlate them with projected radial velocity maps (see Fig.~\ref{fig:visualization} for visualization), enabling a redshift-localized measurement of baryonic feedback. 
We forecast the detectability for two weak lensing survey configurations, Rubin LSST~Y10 and the Roman kinematic-lensing sample \cite{XuEifler2023}, each combined with the CMB experiments the Atacama Cosmology Telescope (ACT), the Simons Observatory (SO), and CMB-HD \citep{ACT:2023kun,SO:ScienceGoalsForecasts,CMBHD:SnowmassWP}. 
As summarized in Fig.~\ref{fig:max_snr}, all configurations reach approximately 5$\sigma$ or higher, and the signal-to-noise ratio rises steeply with CMB sensitivity. 
Translated through the baryonic suppression, this places the sensitivity of this probe sufficient for per cent-level modeling requirement for the next generation weak lensing surveys. 
Experiments like SO can deliver few-percent precision on the matter power spectrum with Rubin LSST~Y10, while CMB-HD pushes both surveys into the sub-percent regime. The same week our paper was submitted, a closely related estimator to probe $P_\mathrm{me}$ appeared in \cite{Hadzhiyska2026}.

The paper is organized as follows. Section~2 introduces the three observables to construct the kSZ template; Section~3 develops intuition for kSZ template construction across multiple redshift bins; Section~4 presents the formalism for cross-correlation; Section~5 presents the SNR forecasts; we conclude in Section~6. 

Throughout this work we assume a flat $\Lambda$CDM cosmology with the \emph{Planck} 2018 best-fit parameters \cite{Planck2020}: $H_0 = 67.66\,\mathrm{km\,s^{-1}\,Mpc^{-1}},\, \omega_{\rm b} = 0.02242,\, \omega_{\rm cdm} = 0.11933,\, \ln(10^{10}A_s) = 3.047,\, n_s = 0.9665,\, \tau_{\rm reio} = 0.0561$, one massive neutrino of mass $m_\nu = 0.06\,\mathrm{eV}$.

\section{Theoretical background}
\label{sec:bkg}
This section provides the theoretical framework for the three observables used in this analysis: kSZ temperature anisotropy, weak lensing convergence, and reconstructed velocity field. In the subsections below, we describe the signal modeling and dominant noise contributions for each probe.

\subsection{kSZ anisotropy}
\label{subsec:ksz}
The kSZ effect is the shift in the energy of the CMB photons caused by Thomson scattering off free electrons moving with a bulk velocity. The kSZ map is a projection of the momentum density field of electrons with the projection kernel $W^{\mathrm{kSZ}}$ given by:
\begin{equation}
    W^{\mathrm{kSZ}}(\chi) = a\bar{n}_\mathrm{e}(a)\sigma_\mathrm{T}\, \mathrm{e}^{-\tau(\chi)}\,,
\end{equation}
where $\sigma_\mathrm{T}$ is the Thomson scattering cross-section, $\tau$ is the integrated optical depth  along the line of sight (LoS), $\bar{n}_\text{e}$ is the homogeneous free electron number density, and $\chi$ is the comoving distance at scale factor $a$.

At linear order in perturbations, the kSZ effect vanishes on small scales due to cancellations between positive and negative velocities along the LOS. Therefore, the kSZ anisotropy on small scales is generated at second order, by the transverse component of the large scale velocity field modulated by the small scale  density perturbations. The kSZ map $T$ can be defined as a function of the angular position $\boldsymbol{\theta}$ in the plane of the sky with respect to some line of sight direction as:
\begin{equation}
    T(\boldsymbol{\theta}) = \int d\chi\, W^{\mathrm{kSZ}}(\chi)\, \delta_\mathrm{e}\left(\chi, \chi \boldsymbol{\theta}, z(\chi)\right)\, v_\mathrm{r} \left(\chi, \chi \boldsymbol{\theta}, z(\chi)\right)\,,
\end{equation}
where $\delta_\mathrm{e}$ and $v_\mathrm{r}$ are the electron density contrast and radial velocity field, respectively.

In practice, the kSZ signal is extracted from the observed CMB temperature maps that contain additional contributions, which together define the effective noise budget. On large scales ($\ell \lesssim 3000$), the primary CMB dominates and acts as the main source of noise. The intermediate scales ($3000 \lesssim\ell \lesssim 5000$) are dominated by foreground contributions from the thermal Sunyaev--Zeldovich effect, the cosmic infrared background, and radio point sources, while on the smallest scales ($\ell \gtrsim 5000$), the instrument detector noise starts becoming important.

\subsection{Weak lensing}
\label{subsec:weaklensing}
The weak lensing map is sensitive to the matter density field projected along the LoS. The lensing convergence field for a galaxy sample $i$ located at a comoving distance $\chi_i$ is given by:
\begin{equation}
    \kappa_i(\boldsymbol{\theta}) = \int_0^{\chi_i} d\chi\, W_i^\kappa(\chi)\, \delta_\mathrm{m}\left(\chi, \chi \boldsymbol{\theta}, z(\chi)\right)\,,
\end{equation}
where $\delta_\mathrm{m}$ is the matter density contrast and the lensing projection kernel $W^i_\kappa$ is given by:
\begin{equation}
    W_i^\kappa(\chi) = \frac{3H_0^2\Omega_\mathrm{m}}{2\mathrm{c}^2}\frac{\chi}{a(\chi)}\int_\chi^{\chi_h}d\chi'\,\frac{dz}{d\chi'}\frac{n_i(z(\chi'))}{\bar{n}_i}\left(1-\frac{\chi}{\chi'}\right)\,,\label{eq:lensingkernel}
\end{equation}
where $\chi_h$ is the comoving distance to the horizon, $n_i(z)$ and $\bar{n}_i$ denote the redshift distribution and the surface density of the galaxy sample. We model the effect of photo-$z$ uncertainty on $n_i$ by convolving the
true redshift distribution of source galaxies $n_\mathrm{true}$ with a Gaussian distribution of standard deviation $\sigma_z(1+z)$ and zero redshift bias,
\begin{align}
n_i(z)= n_\mathrm{true}(z) \int_{z_\mathrm{ph}^{i,\mathrm{min}}}^{z_\mathrm{ph}^{i,\mathrm{max}}}
   dz_\mathrm{ph}\; \frac{1}{\sqrt{2\pi}\,\sigma_z(1+z)}
   \exp\!\left[-\frac{(z_\mathrm{ph}-z)^2}{2\,\sigma_z^2(1+z)^2}\right]\,.
\end{align}
We summarize the weak lensing survey specifications in Table~\ref{tab:wl_surveys}. The source galaxy redshift distribution is described below:
\paragraph{Rubin LSST Year 10 source sample.}
We use the Rubin LSST Year 10 source distribution from the DESC Science Requirements Document \cite{LSST-DESC-SRD_2018}, modeled as a Smail-form distribution,
\begin{equation}
    n_\mathrm{true}(z) \propto z^2 \,\exp\!\left[-(z/z_0)^{\alpha}\right]\,,
    \label{eq:smail}
\end{equation}
with $z_0 = 0.11$ and $\alpha = 0.68$. The distribution has a median redshift $z_\mathrm{med} \simeq 1$ and extends to $z \simeq 4$, with the bulk of the sample concentrated in $0.2 \lesssim z \lesssim 2$.
\paragraph{Roman kinematic-lensing sample.}
For the velocity-reconstruction sample we adopt the Roman Space Telescope kinematic-lensing (KL) galaxy sample of~\cite{XuEifler2023}, a subset of the High Latitude Spectroscopic Survey selected to have at least one of H$\alpha$ or [\textsc{O\,iii}] resolved within the grism spectral range. We take the true redshift distribution directly from Fig.~3 of~\cite{XuEifler2023}; it covers $0.5 \lesssim z \lesssim 3$, with the H$\alpha$ subsample contributing predominantly at $z \lesssim 2$ and [\textsc{O\,iii}] dominating at higher redshifts.

In practice, the weak lensing measurements are limited by the uncertainty in the intrinsic shape and orientation of the galaxies, referred to as shape noise. The corresponding contribution to the convergence power spectrum for source galaxy sample $i$ is given by:
\begin{equation}
  N_\ell^{ij}  = \frac{\sigma_\epsilon^2}{\bar{n}_i}\delta_{ij}\,,
  \label{eq:og_shapenoise}
\end{equation}
where $\sigma_\epsilon$ is the root-mean square intrinsic ellipticity per shear component (see Table~\ref{tab:wl_surveys} for details).

\subsection{Reconstructed velocity}
\label{subsec:recons_vel}
The 3D velocity field can be reconstructed from the galaxy overdensity field using the linearized continuity equation in redshift space. Projecting along the LoS with an appropriate kernel $\mathcal{W}_v$, the 2D radial velocity map in redshift bin $i$ is given by:
\begin{equation}
\mathcal{V}_i(\boldsymbol{\theta}) = \int d\chi\, \mathcal{W}^v_i(\chi)\, v_\mathrm{r}(\chi, \chi\boldsymbol{\theta})\,.
\end{equation}
This reconstruction can be performed using galaxy density fields from both spectroscopic and photometric surveys. Spectroscopic samples yield a more accurate 3D velocity reconstruction due to their precise redshift measurements, while photometric surveys enable tomographic reconstructions on large scales due to larger redshift uncertainties. The primary sources of noise are galaxy shot noise, redshift uncertainties, and nonlinear effects on small scales. 

\section{Intuition and motivation}
The goal of this section is to develop intuition for the information encoded in the cross-correlation signal proposed in this work. We begin by introducing a toy model for the cross-correlation in a single redshift bin and demonstrate how it directly probes the baryonic feedback effects. We then generalize the formalism to multiple redshift bins by employing the nulling technique that enables us to isolate and characterize the redshift evolution of baryonic feedback.

\subsection{Single redshift bin}
We consider the case of a single redshift bin, in which the kSZ template is constructed by taking the product of the convergence field with the reconstructed velocity field in the same redshift bin. For simplicity we drop the lensing and velocity projection kernels and consider an infinitesimally narrow redshift bin. The kSZ template in redshift bin $i$ becomes $\hat{T}_{i,i} \approx \kappa_i \mathcal{V}_i \approx \delta_\mathrm{m} v_\mathrm{r} (z_i)$, and the true kSZ signal in that bin is $T \approx \delta_\mathrm{e}v_\mathrm{r} (z_i)$. The leading order term in the cross-correlation signal is,
\begin{equation}
    \langle T\,\hat{T}_{i,i}\rangle \approx \langle \delta_\mathrm{e}\,\delta_\mathrm{m}\rangle\,\langle v_\mathrm{r}\, v_\mathrm{r}\rangle = P_\mathrm{em} (z_i) \langle v_\mathrm{r}\, v_\mathrm{r}\rangle (z_i)\,.
\end{equation}

To understand the significance of $P_\mathrm{em}$ in modeling the baryonic feedback, we can decompose the total matter power spectrum $P_\mathrm{mm}$ into its dark matter and baryonic components:
\begin{equation}
P_\mathrm{mm} \;=\;
\underbrace{
  \left(\frac{\Omega_\mathrm{c}}{\Omega_\mathrm{m}}\right)^{\!2} P_\mathrm{cc}}_{\textstyle P_\mathrm{cdm}}
\;+\;
\underbrace{
  2\,\left(\frac{\Omega_\mathrm{c}\Omega_\mathrm{b}}{\Omega_\mathrm{m}^{2}}\right)\, P_\mathrm{cb}
  \;+\;
  \left(\frac{\Omega_\mathrm{b}}{\Omega_\mathrm{m}}\right)^{\!2} P_\mathrm{bb}
}_{\textstyle \delta P}\,,
\label{eq:Pmm-split}
\end{equation}
where each term is weighted by the corresponding mass fractions ($\Omega_\mathrm{c}/\Omega_\mathrm{m}\approx 0.87$ for cold dark matter (cdm) and $\Omega_\mathrm{b}/\Omega_\mathrm{m}\approx 0.13$ for baryons). Assuming free electrons trace the baryons, $P_\mathrm{em} \approx P_\mathrm{bm}$, the  baryonic contribution can be expressed as:
\begin{equation}
\begin{aligned}
\frac{\delta P}{P_\mathrm{mm}} 
&=
2
\left(\frac{\Omega_\mathrm{c}\Omega_\mathrm{b}}{\Omega_\mathrm{m}^{2}}\right)
\, 
\frac{P_\mathrm{cb}}{P_\mathrm{mm}}
\;+\;
\left(\frac{\Omega_\mathrm{b}}{\Omega_\mathrm{m}}\right)^{\!2} 
\frac{P_\mathrm{bb}}{P_\mathrm{mm}}
\\
&=
\underbrace{
2\left(
\frac{\Omega_\mathrm{b}}{\Omega_\mathrm{m}}
\right) 
}_{\simeq 26\%}
\frac{P_\mathrm{bm}}{P_\mathrm{mm}}
- 
\underbrace{
\left(
\frac{\Omega_\mathrm{b}}{\Omega_\mathrm{m}}
\right)^2
}_{\simeq 1.7 \%}
\frac{P_\mathrm{bb}}{P_\mathrm{mm}}
.
\end{aligned}
\label{eq:Pmm-split-fractional}
\end{equation}
Thus measuring $P_\mathrm{bm}$ perfectly and neglecting the $P_{bb}$ already enables an accuracy of $1.7\%$ on the matter power spectrum on quasi-linear scales ($k\lesssim0.3\,h\,\mathrm{Mpc}^{-1}$).
This is precisely the goal of our approach, which measures $P_\mathrm{em}$ directly from kSZ and shear, as opposed to measuring $P_\mathrm{eg}$ from kSZ and galaxies.

\subsection{Multiple redshift bins}
The single-bin toy model discussed above is sensitive to the baryonic contribution averaged across the full redshift range probed by the projection kernels. Since the weak lensing kernels are broad, extending from the observer to the source, any redshift dependence of baryonic feedback is lost in the recovered signal. To recover this evolution, we employ the BNT transform \cite{Bernardeau_2014}, a nulling scheme \cite{HutererWhite_2005} which produces a lensing kernel localized in redshift through a linear combination of kernels from three adjacent source-galaxy redshift bins via the relation:
\begin{equation}
    \mathcal{W}^\kappa_i = \sum_{j=i-2}^{i} p_{ij} \, W^\kappa_j\,,\label{eq:nulling1}
\end{equation}
where the coefficients $p_{ij}$ are chosen so that the resulting kernel vanishes for distances smaller than the lowest of the three source distances.  

\begin{figure}[t]
    \centering
    \includegraphics[width=\linewidth]{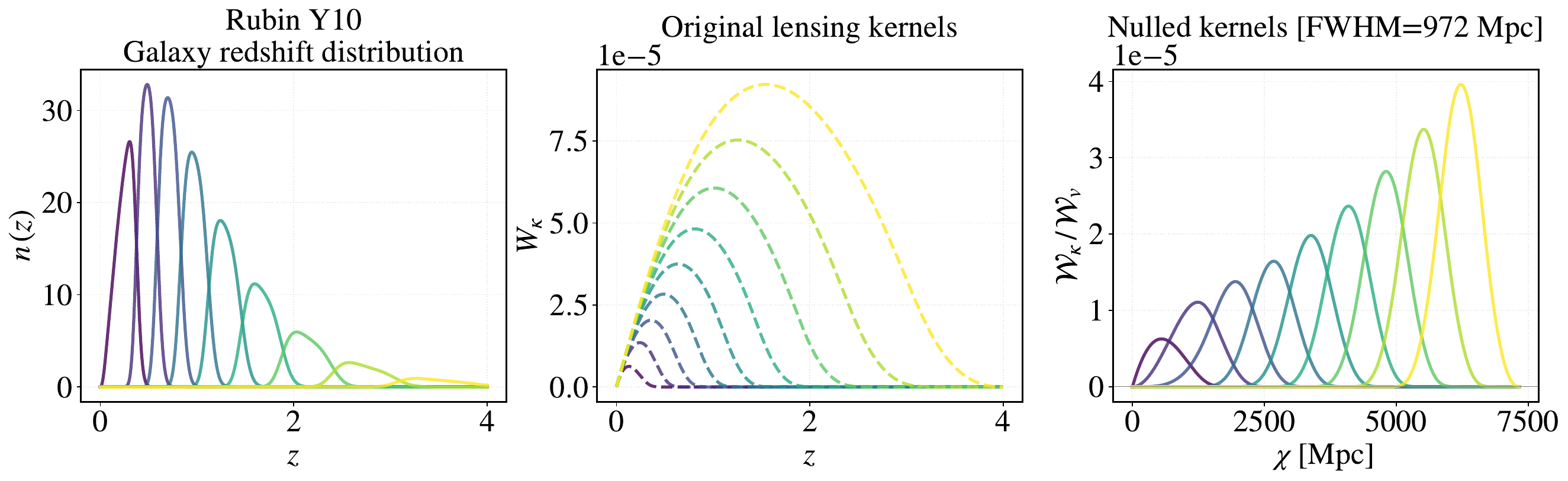}
    \includegraphics[width=\linewidth]{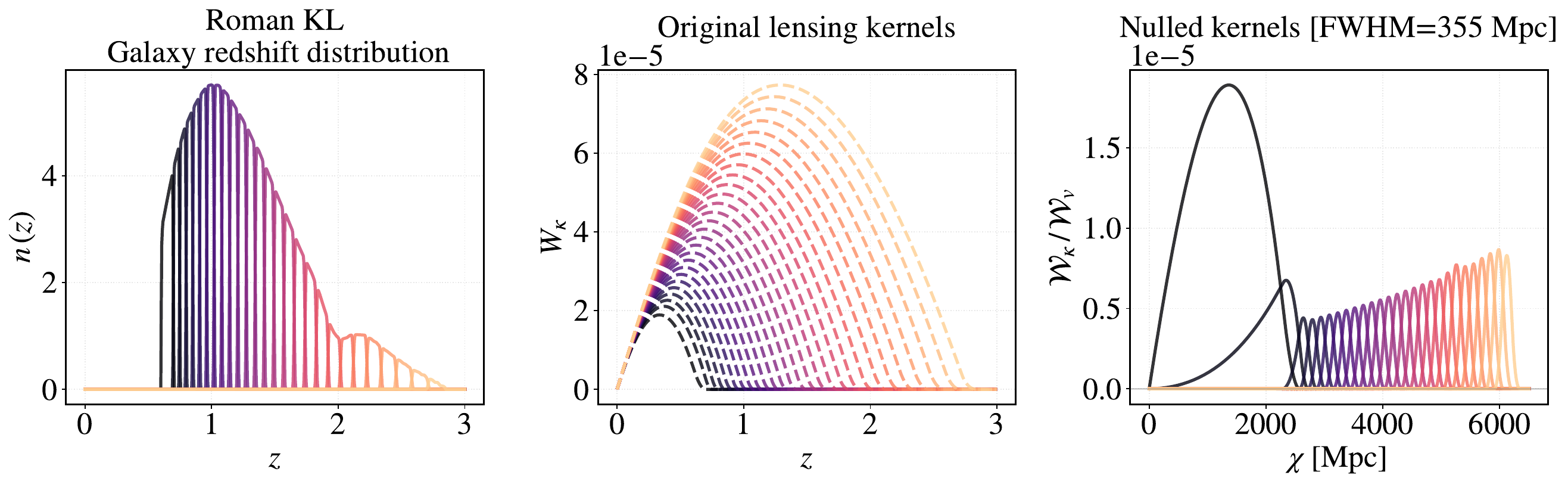}
    \caption{\textit{Top (Bottom)}: Construction of  $N=9$ ($N=20$) nulled lensing kernels of approximately uniform width $\Delta\chi \approx 977$ ($355$) Mpc for Rubin LSST (Roman KL). \textit{Left}: The source galaxy tomographic bins, with the lowest bin of width $\Delta\chi$ and subsequent bins of width $\Delta\chi/2$. \textit{Centre}: The corresponding weak lensing kernels $W^\kappa_i(z)$, each extending from the observer to the source bin. \textit{Right}: The nulled lensing kernels $\mathcal{W}^\kappa_i(\chi)$ localized in comoving distance with the same width.}
    \label{fig:nulling}
\end{figure}

The original lensing kernel given in Eq. \eqref{eq:lensingkernel} is a second-order polynomial in the lens distance $\chi$. The nulling operation imposes two homogeneous conditions, which admit a non-trivial solution  in the presence of at least three original kernels. The remaining one-parameter freedom, corresponding to the overall normalization of the nulled kernel, is fixed by setting  $p_{ii} = 1$. Consequently, nulling is possible only for atleast three original kernels. However, to match the number of nulled kernels to the number of original source bins, we follow the standard convention and define the first two nulled kernels as:
\begin{equation}
    \mathcal{W}^\kappa_1 = W^\kappa_{1}\,, \hspace{0.2cm}
    \mathcal{W}^\kappa_2 = -W^\kappa_{1} + W^\kappa_{2}\,. \label{eq:nulling2}
\end{equation}
As an example, in Fig.~\ref{fig:nulling}, we illustrate the construction of nulled kernels for Rubin LSST Y10 and Roman KL surveys following the procedure outlined above. We present a visual representation of the cross-correlation signal for multiple-bin case in Fig.~\ref{fig:visualization}.

\section{Formalism}
In this section we introduce the notation for the true kSZ signal, the template, and the cross-correlation signal, and derive the final expressions that are used in the forecasts later. 

\subsection{Template kSZ signal}
We construct a template for the kSZ map $\hat{T}$ using the map of the projected matter density field $\mathcal{K}_i$ in redshift bin $i$, and the projected radial velocity field $\mathcal{V}_\alpha$ in redshift bin $\alpha$ as:

\begin{align}
    \hat{T}_{i,\alpha} (\boldsymbol{\theta})
    &= \mathcal{K}_i(\boldsymbol{\theta})\,\mathcal{V}_\alpha(\boldsymbol{\theta})\nonumber\,,\\
    &\approx   
     \int d\chi\, \underbrace{\mathcal{W}_i^{\kappa}(\chi) \,\frac{D(\chi)}{D(\chi_{i})}}_{\textstyle H^\kappa_i(\chi)}\,\delta_\mathrm{m}\left(\chi, \chi\boldsymbol{\theta}, z(\chi_i)\right) \int d\chi'\, \underbrace{\mathcal{W}_\alpha^{v}(\chi')\, \frac{D_v(\chi')}{D_v(\chi'_\alpha)}}_{\textstyle H^{v}_\alpha(\chi)}\,v_\mathrm{r}\left(\chi', \chi'\boldsymbol{\theta}, z(\chi_\alpha)\right)\,, \label{eq:That} 
\end{align}
where we absorb the redshift evolution into the weight function by defining the growth factor $D(\chi)\,(D_v(\chi)=aH(z)f(z)D(\chi)$, where $f$ is the growth rate) with respect to redshift $z_i$\,($z_\alpha$) at which the lensing (velocity) kernel peaks. For simplicity, we assume the density and velocity kernels to have the same functional form $H^{\kappa}_i=H^{v}_\alpha$, for $i=\alpha$.

\subsection{True kSZ signal}
The true kSZ signal represents the projected momentum density field of electrons and is given by:
\begin{equation}
    T(\boldsymbol{\theta}) \approx \int d\chi\, \underbrace{W^{\mathrm{kSZ}}(\chi)\,\frac{D(\chi)D_v(\chi)}{D(\chi_i)D_v\,(\chi_\alpha)}}_{\textstyle H^{\mathrm{kSZ}}(\chi)}\,\delta_\mathrm{e}\left(\chi, \chi \boldsymbol{\theta}, z(\chi_i)\right)\, v_\mathrm{r} \left(\chi, \chi \boldsymbol{\theta}, z(\chi_\alpha)\right)\,,\label{eq:trueksz}
\end{equation}
where we absorb the redshift evolution of electron density and velocity fields into the kSZ weight function similar to Eq. \eqref{eq:That}. 
To have a more accurate estimate of redshift evolution of the electron density and velocity fields, we set the reference redshift for the true kSZ signal with respect to the template kSZ signal in density redshift bin $z_i$ and velocity redshift bin $z_\alpha$.

\subsection{Cross-correlation signal}

Assuming the snapshot approximation at a fixed redshift $z_i$ ($z_\alpha$) for lensing (velocity) kernels, we arrive at the following expression for the cross-correlation signal:
\begin{align}
    \mathcal{C}^{T\hat{T}_{i,\alpha}}_\ell &=
    \langle T\left(\ell\right)\hat{T}_{i,\alpha}\left(-\ell\right)\rangle
    = \frac{1}{\bar{\chi}_{i\alpha}^2}\int \dfrac{dk_\parallel}{2\pi}\, H^\mathrm{kSZ}(-k_\parallel) \int \frac{dK_\parallel}{(2\pi)}\, H^{\kappa}_i(k_\parallel - K_\parallel)\,H^{v}_\alpha(K_\parallel)\,\nonumber\\
    &\hspace{-1cm}P_{\mathrm{em}}\left(\sqrt{\left(k_\parallel - K_\parallel\right)^2 +  \left(\ell/(\bar{\chi}_{i\alpha})\right)^2}, z(\bar{\chi}_{i\alpha})\right)S(K_\parallel, \sigma_\chi)\int \frac{d^2\boldsymbol{K}_\perp}{(2\pi)^2}\,P_{v_\mathrm{r}v_\mathrm{r}}\left(\sqrt{K_\parallel^2 + |\boldsymbol{K}_\perp|^2}, z(\bar{\chi}_{i\alpha})\right)\,,\label{eq:cross}
\end{align}
where $\chi_i$ and $\chi_\alpha$ are the comoving distances at which the density and velocity kernels peak, respectively. Their mean $\bar{\chi}_{i\alpha} = 1/2\,\left(\chi_i + \chi_\alpha\right)$ approximately characterizes the comoving distance where the two kernels overlap most strongly. $P_{v_\mathrm{r}v_\mathrm{r}}$ denotes the radial velocity power spectrum and $P_\mathrm{em}$ is the full non-linear 3D electron-matter power spectrum with the generalized-NFW profile of Battaglia \cite{Battaglia_2016} using the AGN-feedback best-fit parameters. The reconstructed velocity field is associated with a photo-$z$ smearing factor, $S(K_\parallel, \sigma_\chi) = \mathrm{e}^{-(K_\parallel\sigma_\chi)^2/2}$. We provide the detailed derivation of the cross-correlation signal in App.~\ref{app:cross-signal}.

\section{Detectability}
\subsection{Full covariance}
Under the Gaussian approximation, the covariance between $\mathcal{C}^{T\hat{T}_{i,\alpha}}_\ell$ and $\mathcal{C}^{T\hat{T}_{j,\beta}}_{\ell'}$, where $i,j$ label density bins and $\alpha,\beta$ label velocity bins, is given by:
\begin{equation}
    \text{Cov}\left(\mathcal{C}^{T\hat{T}_{i,\alpha}}_\ell,\, \mathcal{C}^{T\hat{T}_{j,\beta}}_{\ell'}\right) = \frac{1}{(2\ell+1)f_\mathrm{sky}}\left(\mathcal{C}^{\mathcal{K}_i \mathcal{K}_j}_{\ell} + \mathcal{N}^{\mathcal{K}_i \mathcal{K}_j}_{\ell}\right)\left(\mathcal{C}^{\mathcal{V_\alpha V_\beta}} + \mathcal{N}^{\mathcal{V_\alpha V_\beta}}\right)\left(\mathcal{C}^{TT}_{\ell} + \mathcal{N}^{TT}_{\ell}\right)\delta_{\ell\ell'}\,, \label{eq:cov}
\end{equation}
where the cosmic variance contributions to the convergence and velocity fields are given by,
\begin{align}
    \mathcal{C}^{\mathcal{K}_i \mathcal{K}_j}_{\ell}
    &=\int d\chi \,\frac{\mathcal{W}^\kappa_i\,(\chi)\,\mathcal{W}^\kappa_j\,(\chi)}{\chi^2}\,P_{\mathrm{mm}}\left(k = \ell/\chi,\,z(\chi)\right)\,,\\
    \mathcal{C}^{\mathcal{V_\alpha V_\beta}}
    &= \int \frac{dK_\parallel}{2\pi}\,H^{v}_\alpha(K_\parallel)\,H^{v}_\beta(-K_\parallel)\,S^2(K_\parallel,\sigma_\chi)\int\frac{d^2\boldsymbol{K}_\perp}{(2\pi)^2}\,P_{v_\mathrm{r}v_\mathrm{r}}\!\left(\sqrt{K_\parallel^2 + |\boldsymbol{K}_\perp|^2},\,z(\bar{\chi}_{\alpha\beta})\right)\,,
\end{align}
where
$f_\mathrm{sky}$ denotes the observed sky fraction and $\mathcal{C}^{TT}_{\ell}$ represents the full CMB power spectrum. The noise terms are the shape noise of the cosmic shear measurement, $\mathcal{N}^{\mathcal{K}_i \mathcal{K}_j}_\ell$; the reconstruction noise of the velocity field, $\mathcal{N}^{\mathcal{V}_\alpha \mathcal{V}_\beta}$; and the combined primary CMB, foreground, and detector noise of the temperature map, $\mathcal{N}^{TT}_\ell$. We ignore the Gaussian contribution from the cross-term between convergence and radial velocity in the covariance matrix and the cross-correlation signal. It has been shown in the literature \cite{Wayland2026} that the cross-term is subdominant compared to the autocorrelation term and becomes even less important in the presence of shape noise, which dominates the cosmic variance terms on small scales. Unlike the original kernels, whose shape noise contributions are uncorrelated between different bins, nulled kernels constructed from three adjacent overlapping source bins inherit correlated shape noise. The shape noise covariance between nulled lensing kernels can be obtained using Eq.~\eqref{eq:og_shapenoise} and Eq.~\eqref{eq:nulling1}:
\begin{equation}
    \mathcal{N}^{\mathcal{K}_i\,\mathcal{K}_j}_\ell = \sum_{k} p_{ik}\,p_{jk}\,\frac{\sigma_\epsilon^2}{\bar{n}_{k}} \,.\label{eq:shapenoise}
\end{equation}
\begin{figure}[t]
    \centering
    \includegraphics[width=0.5\linewidth]{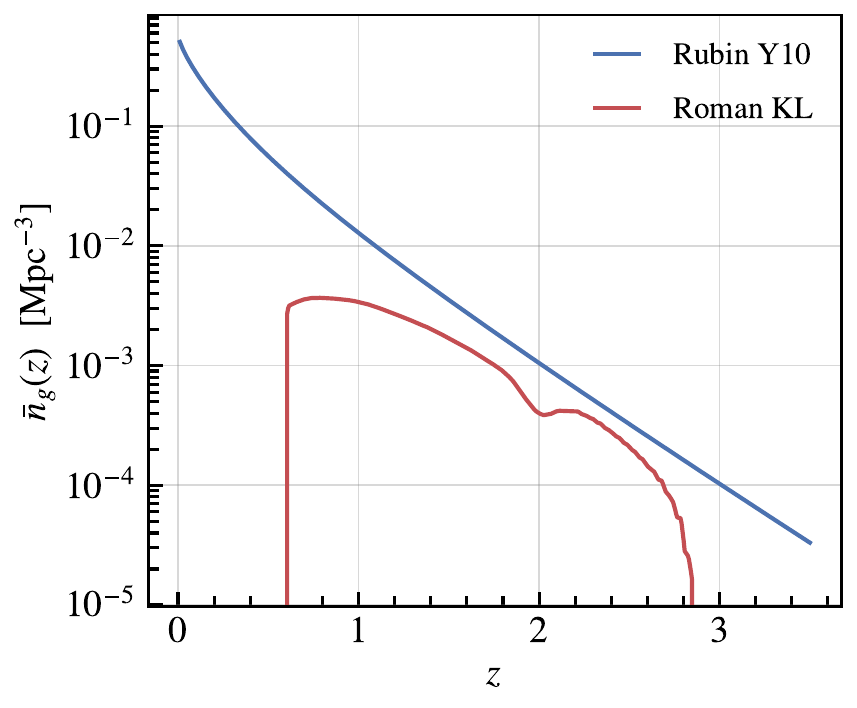}
    \caption{Comoving galaxy number density $\bar{n}_\mathrm{g}(z)$ for the two source samples used in this work: Rubin LSST~Y10 (blue), computed from the Smail-form redshift distribution (see Eq.~\eqref{eq:smail}) from~\cite{LSST-DESC-SRD_2018}, and the Roman~KL sample (red) from~\cite{XuEifler2023}. The Roman~KL sample does not have galaxies at $z \lesssim 0.5$ because its selection requires the detection of H$\alpha$ line inside the grism.}
    \label{fig:ng}
\end{figure}
The noise power spectrum for the velocity field is given by:
\begin{equation}
    \mathcal{N}^{\mathcal{V_\alpha V_\beta}} = \int \frac{dK_\parallel} {2\pi}\,\int\dfrac{d^2\boldsymbol{K}_\perp}{(2\pi)^2}\,\tilde{H}_\alpha^{v}(K_\parallel)\,\tilde{H}_\beta^{v}(-K_\parallel)\frac{K_\parallel^2}{|K|^4}\,\delta_{\alpha\,\beta}\,,\label{eq:vnoise}
\end{equation}
where the modified velocity kernel,
\begin{align}
    \tilde{H}_\alpha^{v}(K_\parallel) 
    &= \mathrm{FT}\!\left[H_\alpha^v(\chi) \sqrt{\frac{1}{\bar{n}_\mathrm{g}(\chi)\,b(\chi)^2}} \right]\,,
\end{align}
is the Fourier transform (FT) of the velocity kernel $H^{v}_\alpha(\chi)$ with the inverse-noise weighting set by the survey's comoving galaxy number density $\bar n_\mathrm{g}$ and linear bias $b$. 
We adopt Rubin LSST Year~10 specifications with galaxy bias $b(z) = 0.95\,\left(D(z=0)/D(z)\right)$~\cite{LSST-DESC-SRD_2018} between $z = 0$ and $z \simeq 3$ and Roman High Latitude Spectroscopic Survey (HLSS) emission line galaxy (ELG) specifications~\cite{Zhai_Wang2021} across $1 \lesssim z \lesssim 3$, with linear bias $b(z) \simeq 0.88\,z + 0.49$ for H$\alpha$ ELGs in $1<z<2$, and $b(z) \simeq 0.98\,z + 0.49$ for [\textsc{O\,iii}] ELGs in $2<z<3$. The comoving number densities for the two surveys are shown in Fig.~\ref{fig:ng}.

\subsection{True-template kSZ correlation coefficient}
\begin{figure}[t]
    \centering
    \includegraphics[width=0.5\linewidth]{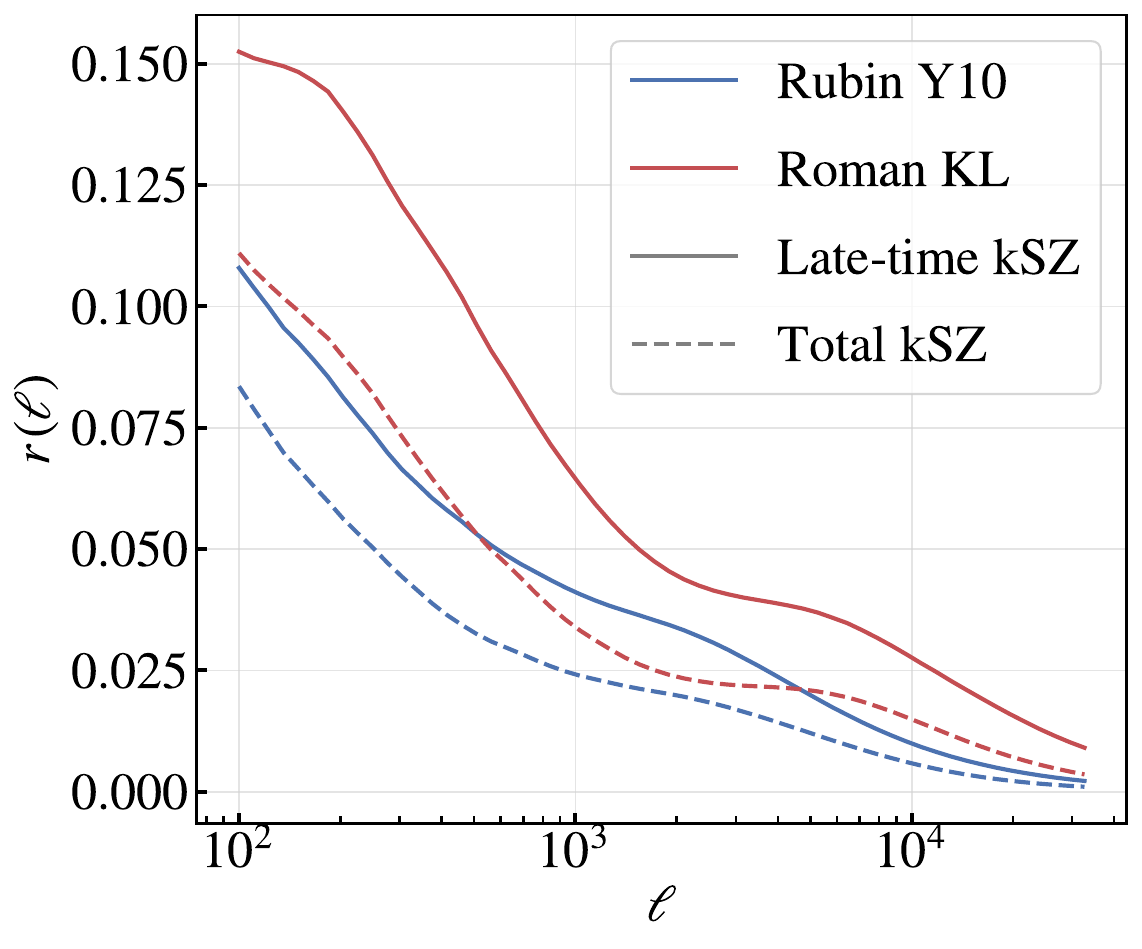}
    \caption{Correlation coefficient $r(\ell)$ between the MSE kSZ template and the true total (dashed lines) and late-time kSZ signal (solid lines) defined in Eq.~\eqref{eq:cross-coeff}, for the Rubin LSST~Y10 (blue) and Roman~KL (red) source samples. $r(\ell)$ evaluated at the tomographic kernel width that maximizes SNR: 977~Mpc for Rubin LSST~Y10 and 355~Mpc for Roman~KL.}
    \label{fig:r_vs_ell_width}
\end{figure}

For a minimum mean squared error (MSE) estimator $\hat{\mathcal{T}}$, the correlation coefficient between the kSZ template and the true-kSZ signal is given by: 
\begin{align}
    \hat{\mathcal{T}}(\ell) &= \mathbf{w}(\ell)^{\top}\,\hat{\boldsymbol{T}}(\ell)\,, \qquad
    \mathbf{w}(\ell) = \mathbf{C}_{\hat{\boldsymbol{T}}\hat{\boldsymbol{T}}}^{-1}(\ell)\,\langle T\,\hat{\boldsymbol{T}}\rangle(\ell)\,, \\[4pt]
    r(\ell) &= \frac{\langle T\,\hat{\mathcal{T}}\rangle(\ell)}{\sqrt{\langle \hat{\mathcal{T}}\,\hat{\mathcal{T}}\rangle(\ell)\,\langle T\,T\rangle(\ell)}} = \sqrt{\frac{\langle T\,\hat{\boldsymbol{T}}\rangle^{\top} (\ell)\,\mathbf{C}_{\hat{\boldsymbol{T}}\hat{\boldsymbol{T}}}^{-1}(\ell)\,\langle T\,\hat{\boldsymbol{T}}\rangle (\ell)}{\langle T\,T\rangle(\ell)}}\,.\label{eq:cross-coeff}
\end{align}
The correlation coefficients of kSZ template with respect to total and late-time kSZ are shown in Fig.~\ref{fig:r_vs_ell_width}. The Roman~KL template achieves a higher correlation than Rubin LSST~Y10 due to its narrower nulled kernels that suffer lower LoS velocity cancellations. 

\subsection{SNR forecasts}
The signal-to-noise-ratio (SNR) is given by,
\begin{equation}
    \mathrm{SNR}^2 = \sum_\ell \boldsymbol{\mathcal{C}}_\ell^{\top}\,\boldsymbol{\Sigma}_\ell^{-1}\,\boldsymbol{\mathcal{C}}_\ell\,,
\end{equation}
where $\boldsymbol{\mathcal{C}}_\ell$ is the data vector obtained by stacking $C_\ell^{T\hat{T}_{i,\alpha}}$ defined in Eq.~\eqref{eq:cross} over all pairs $(i,\alpha)$, and $\boldsymbol{\Sigma}_\ell$ is the corresponding covariance matrix given in Eq.~\eqref{eq:cov}.
\begin{table}[t]
\centering
\begin{tabular}{lcc}
\hline\hline
Parameter & Roman KL & Rubin LSST Y10 \\
\hline
Shape noise per component, $\sigma_\epsilon$        & 0.025  & 0.26   \\
Galaxy surface density, $\bar{n}$ [arcmin$^{-2}$] & 4      & 27     \\
Photo-$z$ uncertainty, $\sigma_z$  & 0.002      & 0.03     \\
Survey area [deg$^2$]                                & 2000 & 18000 \\
Sky fraction, $f_{\rm sky}$                                        & 0.048 & 0.44 \\
\hline
\end{tabular}
\caption{Galaxy survey specifications for Rubin LSST~Y10~\cite{LSST-DESC-SRD_2018} and Roman kinematic lensing~\cite{XuEifler2023}.}
\label{tab:wl_surveys}
\end{table}
\begin{figure}[t]
    \centering
    \includegraphics[width=0.45\linewidth]{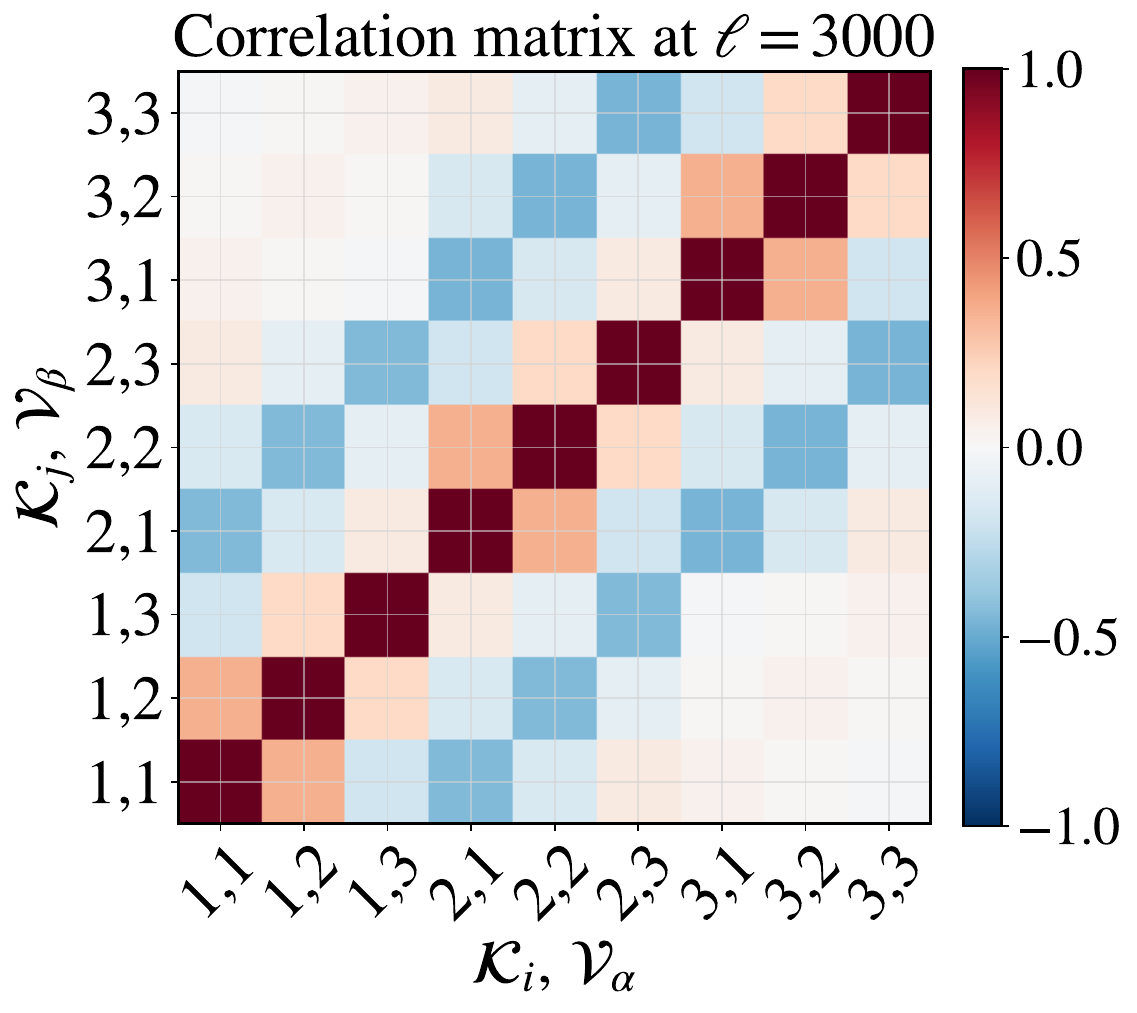}
    \includegraphics[width=0.43\linewidth]{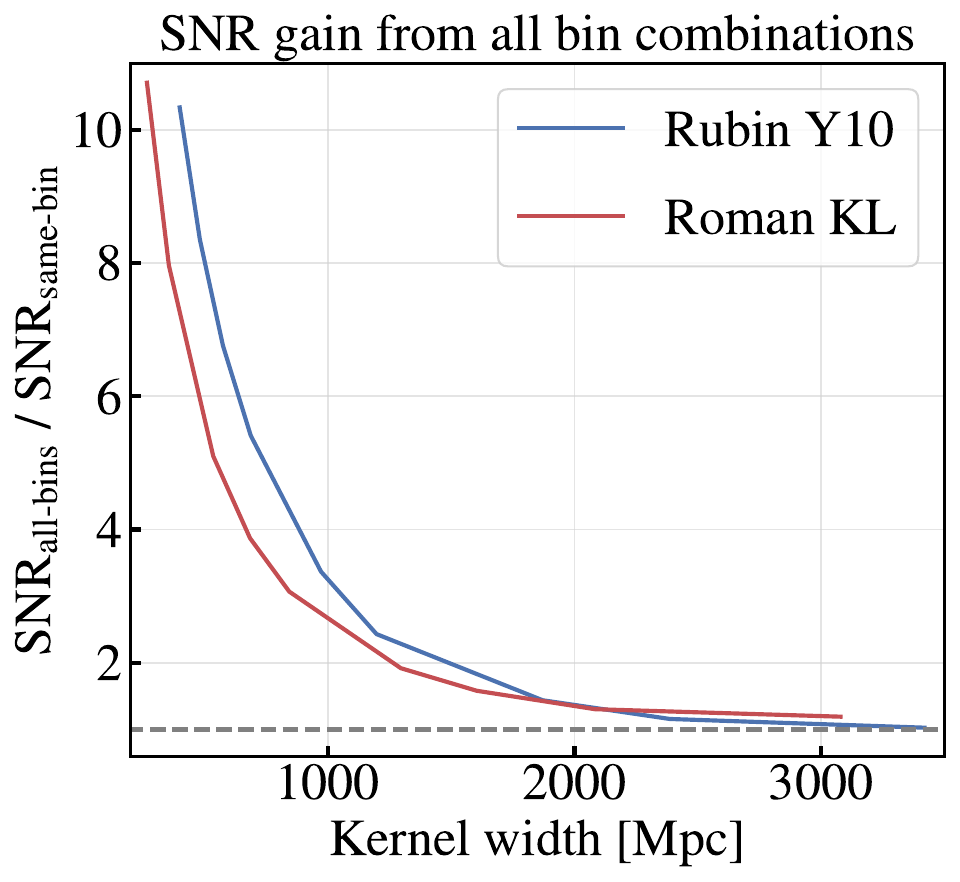}
    \caption{\emph{Left:} Correlation matrix of the kSZ template cross-spectra at $\ell = 3000$, for all redshift-bin pairs ($\mathcal{K}_i,\mathcal{V}_\alpha$). The off-diagonal entries arise because each nulled kernel is constructed as a linear combination of three original tomographic kernels: distinct nulled bins therefore share original kernels, and hence their shape noise, producing correlations between bin pairs. \emph{Right:} Ratio of SNR computed with the full covariance (all bin pairs) to that using the same redshift density and velocity bins, as a function of nulled-kernel width. Accounting for these off-diagonal correlations raises the SNR at all widths, by a factor $\gtrsim\!10$ toward narrower kernels.}
    \label{fig:snr_crossbins_vs_width}
\end{figure}
\begin{figure}[t]
    \centering
\includegraphics[width=0.65\linewidth]{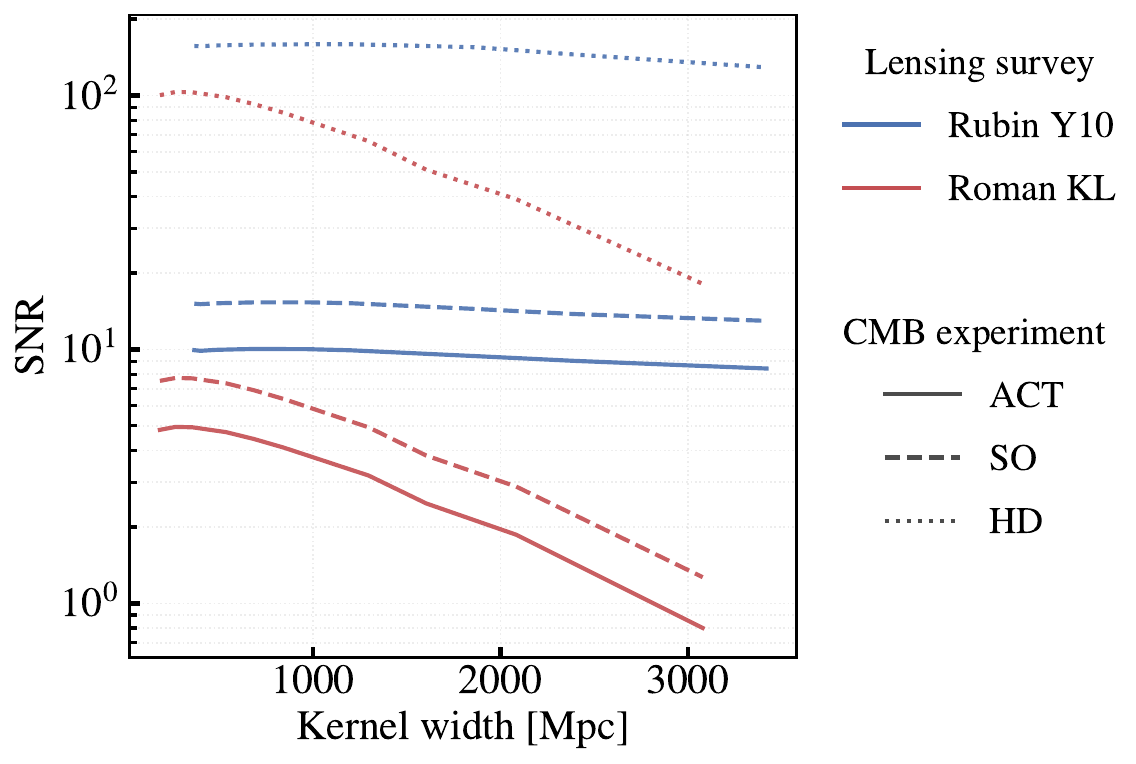}
    \caption{Forecast SNR for the kSZ template cross-correlation as a function of the width of the nulled tomographic kernels for all combinations of weak-lensing and CMB experiments. The Roman KL curves fall off more steeply toward large kernel widths because of the lack of low-redshift galaxies available for velocity reconstruction. The SNR increases strongly for more sensitive CMB experiments.}
    \label{fig:snr_vs_width}
\end{figure}
\begin{table}[t]
\centering
\begin{tabular}{lcc}
\hline\hline
 & Roman KL (355 Mpc) & Rubin LSST Y10 (977 Mpc) \\
\hline
ACT     & 4.96 & 10.05 \\
SO      & 7.72 & 15.38 \\
CMB-HD  & 103.26 & 159.22 \\
\hline\hline
\end{tabular}
\caption{Signal-to-noise ratio (SNR) on the cross-correlation for all combinations of CMB experiments and weak lensing surveys.}
\label{tab:snr_combinations}
\end{table}
\begin{figure}[t]
    \centering
    \includegraphics[width=0.9\linewidth]{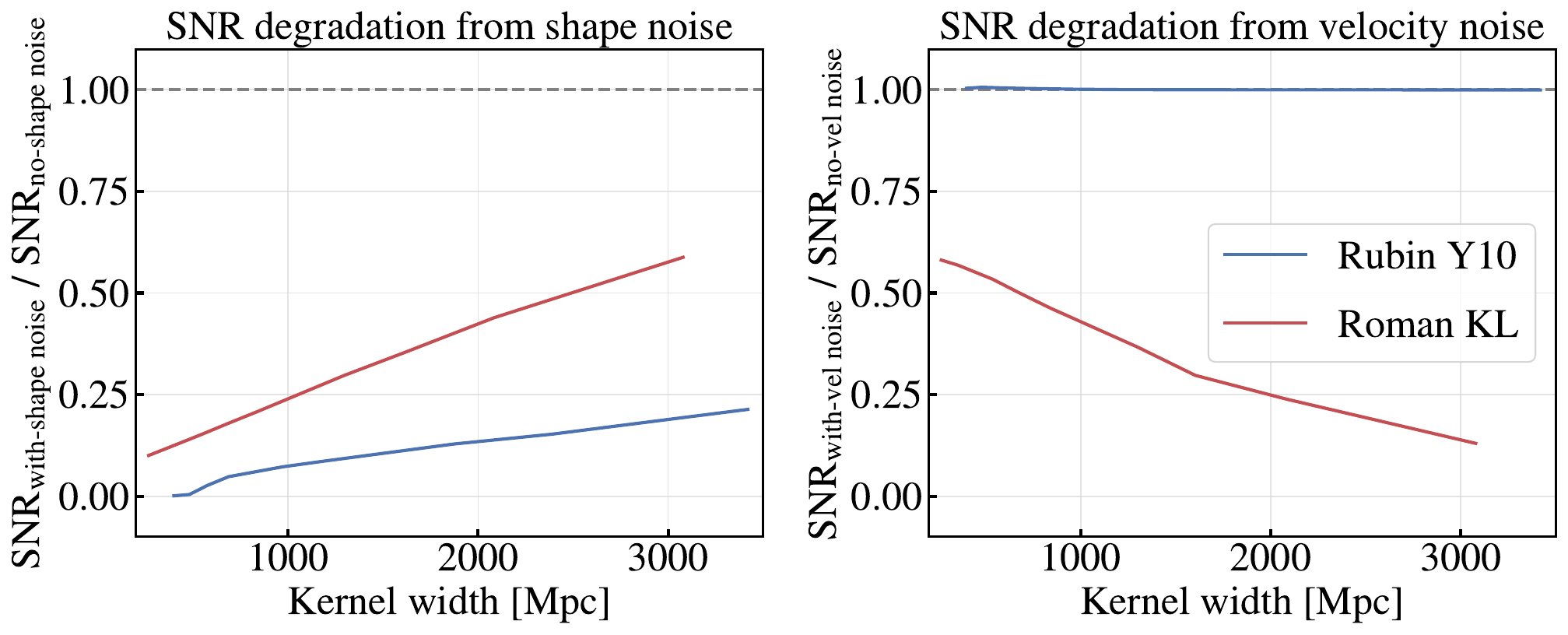}
    \caption{\emph{Left:} Ratio of SNR computed with shape noise to that without shape noise in the absence of velocity reconstruction noise, as a function of nulled-kernel width. Roman KL with better shape noise estimate than Rubin LSST sees a lower degradation.  \emph{Right:} Ratio of SNR computed with velocity reconstruction noise to that without velocity reconstruction noise in the presence of shape noise, as a function of nulled-kernel width. The absence of Roman spectroscopic galaxies at $z\lesssim 0.5$ causes SNR to drop sharply at large widths.}
\label{fig:snr_noise_vs_width}
\end{figure}
The SNR is computed using all combinations of nulled lensing and velocity kernels for Roman KL and Rubin LSST~Y10 with specifications listed in Table~\ref{tab:wl_surveys}. The off-diagonal pairs with non-overlapping redshift support for density and velocity bins contribute negligibly to the cross-correlation signal itself. However, the nulling operation (see equation~\eqref{eq:nulling1}) induces  correlated shape noise between adjacent nulled lensing kernels, visible in the left panel of Fig.~\ref{fig:snr_crossbins_vs_width} as negative secondary diagonals (blue) linking different density bins that share the same velocity bin. Retaining these off-diagonal terms in the SNR calculation (right panel) yields a substantial gain at narrow kernel widths. Narrower kernel widths correspond to larger number of nulled kernels resulting in more off-diagonal pairs contributing to the covariance matrix (see App.~\ref{app:toy-crosscov} for a minimal toy example).

We present the SNR estimates for the cross-correlation measurements for different combinations of the two weak lensing surveys (Roman KL and Rubin LSST~Y 10) and three CMB experiments (CMB-HD, SO, ACT) in Fig.~\ref{fig:snr_vs_width} and Table~\ref{tab:snr_combinations}. 
Each curve exhibits an optimum in kernel width, located at roughly 355~Mpc for Roman KL and 977~Mpc for Rubin LSST weak lensing. The corresponding signal and correlation matrix for all bin combinations are provided in App.~\ref{app:snr_plots}.
All survey combinations reach approximately $5\sigma$ or higher implying that the cross-correlation would be detectable with the forthcoming data.

We show the degradation in SNR due to shape noise and velocity reconstruction noise in Fig.~\ref{fig:snr_noise_vs_width}. Roman KL sample with much lower shape noise per galaxy sees a lower degradation compared to Rubin LSST Y10. For narrower lensing kernels, the number of tomographic bins required for nulling is higher. As a result, the shape noise increasingly degrades the SNR at lower kernel widths. For more details, we refer the reader to a minimal toy model in App.~\ref{app:toy}. We find that the velocity reconstruction noise contribution matters the most at the scale corresponding to the width of the radial velocity kernel (see App.~\ref{app:recon_vel} for details). Since the optimum kernel width for Rubin is on large, linear scales $\approx 977$ Mpc, even in the presence of photo-$z$ uncertainty, the velocity reconstruction is perfect and does not lower the SNR. Since Roman KL does not observe galaxies at $z\lesssim 0.5$ (see Fig.~\ref{fig:ng}), the degradation in SNR due to velocity reconstruction noise is the highest for Roman KL. We expect that the Roman KL SNR estimates would improve if other surveys like Rubin LSST or Dark Energy Spectroscopic Instrument (DESI) are used for velocity reconstruction at lower redshifts.

\section{Conclusion}
We summarize our main results below:
\begin{itemize}
    \item We propose a new cross-correlation estimator, a kSZ template, built by taking a product of tomographic weak lensing convergence and  radial velocity maps reconstructed from galaxy positions. When cross-correlated with the CMB temperature map, the estimator is directly sensitive to the electron--matter power spectrum $P_{\rm em}$, and therefore to the baryonic contribution to the matter power spectrum.
    \item Because this method directly measures the gas distribution around shear fluctuations rather than galaxies, it avoids a key limiting systematic in existing joint cosmic shear--kSZ analyses: the challenge of modeling the galaxy--matter connection and extrapolating towards mass and redshift ranges not sampled by the galaxies. By trading off statistical power for systematic control, this approach provides a powerful independent handle on the baryonic uncertainties in cosmic shear.
    \item 
    Our forecasts are encouraging. First, a detection is imminent (SNR=5 for Rubin$\times$ACT and 10 for Roman$\times$ACT), then the SNR will be good enough to address the baryon issue in Rubin (8 for Rubin$\times$SO and 15 for Roman$\times$SO). Finally, in the future, the SNR should enable highly precise measurements of the electron-matter power spectrum (103 for Rubin$\times$CMB-HD and 160 for Roman$\times$CMB-HD).
    \item The BNT nulling transform provides an intuitive way to localize the kernels in redshift, in order to resolve the line-of-sight Fourier modes required for kSZ. 
    Using the identical equi-width projection kernel for matter field and LoS velocities, we find the optimum kernel widths to be $\sim 977$ Mpc for Rubin~Y10 and $\sim 355$ Mpc for Roman~KL. We explore numerically and intuitively with a toy model the impact of kernel width and photo-z errors on the estimator. However, since this is an invertible linear transformation, the same information is obtained by jointly analyzing the cross-correlations of shear $\times$ velocity $\times$ CMB temperature for all tomographic bins.
    Thus the proposed method does not rely crucially on the nulling transform.
    \item The correlation coefficient between our kSZ template and the true late-time kSZ signal reaches $r \sim {\rm 0.06}$ ($r \sim {\rm 0.04}$) at $\ell \sim 1000$ for Roman KL (Rubin Y10) and decreases towards higher $\ell$. This approach alone is thus not sufficient for applications that seek to clean out CMB maps by removing their kSZ component.
    \item Shape noise is the biggest limiting factor for Rubin~Y10, while velocity-reconstruction noise dominates for Roman~KL, driven by the absence of Roman spectroscopic galaxies at $z \lesssim 0.5$ as shown in Fig.~\ref{fig:snr_noise_vs_width}. Substantial gains are possible if shape noise can be reduced for Rubin. Using an external sample, such as Rubin or DESI, for velocity reconstruction could recover the missing low-redshift information in the Roman KL sample. We leave the development of a fully optimized estimator, including such cross-survey combinations for velocity reconstruction, to future work.
\end{itemize}

\acknowledgments

AG, EK, and TE were supported in part by the Roman
Project Infrastructure Team “Maximizing Cosmological Science with the Roman High Latitude Imaging Survey" (NASA
contracts 80NM0018D0004-80NM0024F0012) and Department of Energy grant DE-SC0025993. 
Work done at SLAC National Accelerator Laboratory was supported by the Department of Energy under contract DE-AC02-76SF00515.
This work was partially supported by the Laboratory Directed Research and Development (LDRD) program at SLAC National Accelerator Laboratory, Project ID 25-012.

\newpage
\appendix
\section{True and template cross-correlation and noise}
\label{app:cross-signal}

We compute the cross-correlation between true kSZ map and template kSZ map in Fourier space. We use the linear growth factor to model the redshift evolution of density and velocity perturbations within a given redshift bin with respect to the kernel peak redshift. This assumption may not be accurate on non-linear scales if the redshift bin is very wide.
\subsection{Non-linear kSZ map in Fourier space}
\label{app:ksz}
To Fourier transform the true kSZ map defined in Eq.~\eqref{eq:trueksz}, we first transform along the line of sight, from the comoving distance $\chi$ to the conjugate mode $k_\parallel$, and then transform the angular coordinate $\boldsymbol{\theta}$ to multipole moment $\boldsymbol{\ell}$. The density power spectrum snapshot is taken at $z(\chi_i)$ and the velocity power spectrum snapshot is taken at $z(\chi_\alpha)$, subject to the peak redshift of the $\kappa$ and reconstructed radial velocity $v_\mathrm{r}$ map. The Fourier transform of the kSZ map $T(\boldsymbol{\ell})$ is given by:
\begin{align}
T(\boldsymbol{k}_\perp) 
&= \int d^2\boldsymbol{x}_\perp\,e^{-i \boldsymbol{k_\perp}\cdot\boldsymbol{x}_\perp}\,T(\boldsymbol{\theta}=\boldsymbol{x}_\perp/\chi)\,,\nonumber\\
&=\int \frac{d^2\boldsymbol{K}_\perp}{(2\pi)^2}\int \frac{d^2\boldsymbol{K}'_\perp}{(2\pi)^2} \int d^2\boldsymbol{x}_\perp\,e^{i( \boldsymbol{K}_\perp+\boldsymbol{K}'_\perp-\boldsymbol{k}_\perp)\cdot\boldsymbol{x}_\perp} \int d\chi\, e^{i(k_\parallel + K_\parallel + K'_\parallel)\chi} \nonumber\\
&\hspace{0.1cm}\int \frac{dk_\parallel}{2\pi} \,H^{\mathrm{kSZ}}(k_\parallel)\int\dfrac{dK_\parallel}{2\pi} \int\frac{dK'_\parallel}{2\pi}\,
\delta_\mathrm{e}(K_\parallel, \boldsymbol{K}_\perp, z(\chi_i))\, v_\mathrm{r} (K'_\parallel, \boldsymbol{K}'_\perp, z(\chi_\alpha))\,.
\end{align}
Integrating over $\chi$ results in a delta function $\delta (k_\parallel + K_\parallel + K'_\parallel)$ which fixes the LoS modes $K'_\parallel = -k_\parallel - K_\parallel$ once $K'_\parallel$ is integrated out. Similarly, integrating over $\boldsymbol{x}_\perp$, fixes $\boldsymbol{K}_\perp' = \boldsymbol{k}_\perp - \boldsymbol{K}_\perp$. After doing a change of variable from $k_\parallel$ to $-k_\parallel$, and from $\boldsymbol{K} \rightarrow \boldsymbol{k} - \boldsymbol{K}$, we get,
\begin{equation}
  T(\boldsymbol{k}_\perp') = \int \frac{dk_\parallel} {2\pi}\,H^{\mathrm{kSZ}}(-k_\parallel)\, \int\frac{d^3\boldsymbol{K}}{(2\pi)^3} \, \delta_\mathrm{e}(k_\parallel - K_\parallel, \boldsymbol{k}_\perp' - \boldsymbol{K}_\perp)\, v_\mathrm{r} (K_\parallel, \boldsymbol{K}_\perp).
\end{equation}
The true kSZ map $T$ is a projection of the product of density and velocity fields, which in Fourier space is a convolution of both the radial and the transverse wavelength modes.

\subsection{Our kSZ template map}
Following the procedure similar to the kSZ signal in \ref{app:ksz}, the Fourier transform of $\hat{T}$ is given by,
\begin{equation}
  \hat{T}_{i,\alpha}(\boldsymbol{k}_\perp') = \int \dfrac{dk'_\parallel} {2\pi}\,H^{\kappa}(-k'_\parallel)\, \int \dfrac{dk''_\parallel} {2\pi}\,H^{v}(-k''_\parallel)\, \int\dfrac{d^2\boldsymbol{K}'_\perp}{(2\pi)^2} \, \delta_\mathrm{m}(k'_\parallel, \boldsymbol{k}_\perp' - \boldsymbol{K}'_\perp)\, v_\mathrm{r} (k''_\parallel, \boldsymbol{K}'_\perp). 
\end{equation}
Our kSZ template $\hat{T}$ is a product of the projected density and projected velocity fields, which in Fourier space is a convolution of only the wavelength modes transverse to the LoS.

\subsection{Cross-correlation signal}
\begin{align}
    \langle T(\boldsymbol{k}_\perp)\,\hat{T}_{i,\alpha}(\boldsymbol{k}'_\perp)\rangle
    &= \int \dfrac{dk_\parallel} {2\pi}\,H^{\mathrm{kSZ}}(-k_\parallel)\,  \int \dfrac{dk'_\parallel} {2\pi}\,H^{\kappa}(-k'_\parallel)\, \int \dfrac{dk''_\parallel} {2\pi}\,H^{v_r}(-k''_\parallel)\,\int\frac{d^3\boldsymbol{K}}{(2\pi)^3} \, \int\dfrac{d^2\boldsymbol{K}'_\perp}{(2\pi)^2}\,\nonumber\\
    &\langle \delta_\mathrm{e}(k_\parallel - K_\parallel, \boldsymbol{k}_\perp - \boldsymbol{K}_\perp)\,\delta_\mathrm{m}(k'_\parallel, \boldsymbol{k}'_\perp - \boldsymbol{K}'_\perp)\,\rangle \, \langle v_r (K_\parallel, \boldsymbol{K}_\perp)\, v_\mathrm{r} (k''_\parallel, \boldsymbol{K}'_\perp)\rangle\,+\,\nonumber\\
    &\langle \delta_\mathrm{e}(k_\parallel - K_\parallel, \boldsymbol{k}_\perp - \boldsymbol{K}_\perp)\,v_r (k''_\parallel, \boldsymbol{K}'_\perp)\rangle \, \langle \delta_\mathrm{m}(k'_\parallel, \boldsymbol{k}'_\perp - \boldsymbol{K}'_\perp)\, \, v_\mathrm{r} (K_\parallel, \boldsymbol{K}_\perp)\rangle\,+\,\nonumber\\
    &\langle \delta_\mathrm{e}(k_\parallel - K_\parallel, \boldsymbol{k}_\perp - \boldsymbol{K}_\perp)\,v_r (K_\parallel, \boldsymbol{K}_\perp)\rangle \, \langle \delta_\mathrm{m}(k'_\parallel, \boldsymbol{k}'_\perp - \boldsymbol{K}'_\perp)\, \, v_\mathrm{r} (k''_\parallel, \boldsymbol{K}'_\perp)\rangle\,. 
\end{align}
The ensemble averages $\langle \delta_\mathrm{e} \delta_\mathrm{m} \rangle$  and $\langle v_\mathrm{r} v_\mathrm{r} \rangle$ pick out $k'_\parallel = -k_\parallel + K_\parallel$, $k''_\parallel=-K_\parallel$, and $\boldsymbol{K}'_\perp=-\boldsymbol{K}_\perp$. The third term only contributes to zero mode $\boldsymbol{k}_\parallel=0$,  $\boldsymbol{k}_\perp=0$ and is zero for finite $\boldsymbol{k}_\perp$. Since we are interested in small scales where $\boldsymbol{k}_\perp \gg \boldsymbol{K}_\perp$, the second term which involves cross power spectrum $\langle \delta_\mathrm{e} v_\mathrm{r} \rangle$, is suppressed as, $P_{\mathrm{e}v_\mathrm{r}} \sim 1/k\,P_{\mathrm{em}}$. As a good approximation, similar to the kSZ power spectrum, the leading order term is given by,
\begin{align}
    \langle T(\boldsymbol{k}_\perp)\,\hat{T}_{i,\alpha}(-\boldsymbol{k}_\perp)\rangle
    &\approx \int \dfrac{dk_\parallel}{2\pi}\, H^\mathrm{kSZ}(-k_\parallel) \int \frac{d^3\boldsymbol{K}}{(2\pi)^3}\, H^{\kappa}(k_\parallel - K_\parallel)\,H^{v}(K_\parallel)\,\nonumber\\
    &P_\mathrm{em}\left(\sqrt{\left(k_\parallel - K_\parallel\right)^2+ \boldsymbol{k}_\perp^2}\right)\,P_{v_\mathrm{r}v_\mathrm{r}}\left(\sqrt{K_\parallel^2 + \boldsymbol{K}_\perp^2}\right)\,.
\end{align}
The cross-correlation peaks in the redshift slice where lensing and velocity kernels overlap. 
At small scales, we can take the Limber approximation which relates the amplitude of transverse Fourier mode $\boldsymbol{k}_\perp$ to the multipole moment $\ell$ by:  $|\boldsymbol{k}_\perp| \approx (\ell+1/2)/\bar{\chi}_{i\alpha}$, where we approximate $\bar{\chi}_{i\alpha}$ to be the comoving distance at which the overlapping region between lensing and velocity kernels peaks. The cross-correlation signal is given by,
\begin{align}
    \langle T(\ell)\hat{T}_{i,\alpha}(-\ell)\rangle
    &= \frac{1}{\bar{\chi}_{i\alpha}^2}\int \dfrac{dk_\parallel}{2\pi}\, H^\mathrm{kSZ}(-k_\parallel) \int \frac{d^3\boldsymbol{K}}{(2\pi)^3}\, H^{\kappa}(k_\parallel - K_\parallel)\,H^{v}(K_\parallel)\,\nonumber\\
    &P_\mathrm{em}\left(\sqrt{\left(k_\parallel - K_\parallel\right)^2+ \left(\frac{\ell+1/2}{\bar{\chi}_{i\alpha}}\right)^2}\right)\,P_{v_\mathrm{r}v_\mathrm{r}}\left(|\boldsymbol{K}|\right)\,.\label{eq:cross1}
\end{align}
To simplify the numerical integration, we use the shift property of the Fourier transform, 
\begin{align}
    \tilde{f}(k) 
    &= \int dx\, e^{-ikx} f(x) = e^{-ikx_0}\,\int dx'\, e^{-ikx'} f(x'+x_0) = e^{-ikx_0}\,\int dx'\, e^{-ikx'} f_\text{shift}(x')\,,
\end{align}
where a change of variables from $x \rightarrow x' = x-x_0$ allows us to separate the oscillatory part of the Fourier transform. For example, if $f(x)$ is a Gaussian centered at $x_0$, the oscillatory part is entirely captured by the exponential phase. 

Applying the shift property to $H^\kappa$, $H^v$, and $H^\mathrm{kSZ}$ kernels which peak at $\chi_i$, $\chi_\alpha$, and $\chi_\mathrm{kSZ}$ given in \eqref{eq:cross1}, we get,
\begin{align}
    \langle T(\ell)\,\hat{T}_{i,\alpha}(-\ell)\rangle
    & = \frac{1}{\bar{\chi}_{i\alpha}^2}\int \dfrac{dk_\parallel}{2\pi}\, \mathrm{e}^{-ik_\parallel (\chi_i - \chi_\text{kSZ})} H_\text{shift}^\mathrm{kSZ}(-k_\parallel) \, \int \frac{d^3\textbf{K}}{(2\pi)^3} \,\mathrm{e}^{-iK_\parallel (\chi_\alpha - \chi_i)}\, H_\text{shift}^{\kappa}(k_\parallel - K_\parallel)\,\nonumber \\
    &\hspace{2cm} H_\text{shift}^{v}(K_\parallel)\, P_\mathrm{em}\left(\sqrt{\left(k_\parallel - K_\parallel\right)^2+ \left(\frac{\ell+1/2}{\bar{\chi}_{i\alpha}}\right)^2}\right)\,P_{v_\mathrm{r}v_\mathrm{r}}(|\textbf{K}|)\,,
    \nonumber\\
   &= \frac{1}{\bar{\chi}_{i\alpha}^2}\int \dfrac{dk_\parallel}{2\pi}\, \mathrm{e}^{-ik_\parallel (\chi_i - \chi_\text{kSZ})} H_\text{shift}^\mathrm{kSZ}(-k_\parallel) \, \mathcal{I}(k_\parallel, \ell)\,.
\end{align}
Using the property that convolution of two functions in Fourier space is the Fourier transform of their product in real space and vice versa, we get,
\begin{align}
   \mathcal{I}(k_\parallel, \ell) &\approx \int \frac{dK_\parallel}{2\pi}\, P_\mathrm{em}(k=\ell/\bar{\chi}_{i\alpha}) \underbrace{H^{\kappa}(k_\parallel - K_\parallel)}_{\textstyle f(k_\parallel - K_\parallel)}\,\underbrace{H^{v}(K_\parallel)\,C_{v_\mathrm{r}v_\mathrm{r}}(K_\parallel)}_{\textstyle g(K_\parallel)}= \mathrm{FT}\left[f(\chi) * g(\chi)\right]\,,\nonumber\\
  f(\chi) &= \int d\chi' H^{\kappa}(\chi')\,\hspace{0.4cm} \mathrm{and} \hspace{0.4cm} g(\chi) = \int d\chi' H^{v}(\chi')\,\xi_{v_\mathrm{r}v_\mathrm{r}}(\chi-\chi')\,. \label{eq:velsmear}
\end{align}
where $H_\text{shift}^{\kappa}$, $H_\text{shift}^{v}$, and $H_\text{shift}^\mathrm{kSZ}$ refer to the shifted lensing, velocity, and kSZ kernels such that they now peak at $\chi=0$, and $\xi_{v_\mathrm{r}v_\mathrm{r}}$ refers to the radial velocity correlation function. The second term in Eq.~\eqref{eq:velsmear} shows that the velocity kernel gets smeared by the radial velocity correlation  function which is positive and falls steeply till $\lesssim 100$ Mpc and then becomes negative and slowly decays to zero over $\sim 1000$ Mpc.

\subsection{Template kSZ auto-correlation without noise}
\begin{align}
    \langle \hat{T}_{i,\alpha}(\boldsymbol{k}_\perp)\hat{T}_{i,\alpha}(\boldsymbol{k}'_\perp)\rangle
    &= \int \frac{dk_\parallel}{2\pi}\int \frac{dk'_\parallel} {2\pi}\int \frac{dk''_\parallel} {2\pi}\int \frac{dk'''_\parallel} {2\pi}\int\dfrac{d^2\boldsymbol{K}_\perp}{(2\pi)^2}\int\frac{d^2\boldsymbol{K}'_\perp}{(2\pi)^2}\,H^{\kappa}(-k_\parallel)\,H^{\kappa}(-k''_\parallel)\,\nonumber\\
    &\hspace{-1cm}H^{v}(-k'_\parallel)\,H^{v}(-k'''_\parallel)
    \langle \delta_\mathrm{m}(k_\parallel, \boldsymbol{k}_\perp - \boldsymbol{K}_\perp)\,\delta_\mathrm{m}(k''_\parallel, \boldsymbol{k}'_\perp - \boldsymbol{K}'_\perp)\,\rangle \, \langle v_\mathrm{r} (k'_\parallel, \boldsymbol{K}_\perp)\, v_\mathrm{r} (k'''_\parallel, \boldsymbol{K}'_\perp)\rangle\,. 
\end{align}
The ensemble averages $\langle \delta_\mathrm{m} \delta_\mathrm{m} \rangle$  and $\langle v_\mathrm{r} v_\mathrm{r} \rangle$ pick out $k''_\parallel = -k_\parallel$, $k'''_\parallel = -k'_\parallel$, $\boldsymbol{K}'_\perp=-\boldsymbol{K}_\perp$, and $\boldsymbol{k}'_\perp=-\boldsymbol{k}_\perp$.
\begin{align}
    \langle \hat{T}_{i,\alpha}(\boldsymbol{k}_\perp)\hat{T}_{i,\alpha}(-\boldsymbol{k}_\perp)\rangle
    &= \int \dfrac{dk_\parallel} {2\pi}\,H^{\kappa}(k_\parallel)\,H^{\kappa}(-k_\parallel)\, \int \dfrac{dk'_\parallel} {2\pi}\,H^{v}(k'_\parallel)\,H^{v}(-k'_\parallel)\,\nonumber\\
    &\int\dfrac{d^2\boldsymbol{K}_\perp}{(2\pi)^2}\,P_\mathrm{mm}(k_\parallel, \boldsymbol{k}_\perp - \boldsymbol{K}_\perp)\,P_{v_\mathrm{r}v_\mathrm
    {r}}(k'_\parallel, \boldsymbol{K}_\perp)\,.
\end{align}
On small scales we can approximate $\boldsymbol{k}_\perp \gg \boldsymbol{K}_\perp$, and use the Limber approximation to set $|\boldsymbol{k}_\perp| \approx (\ell+1/2)/\bar{\chi}_{i}$,
\begin{align}
     \langle \hat{T}_{i,\alpha}(\ell)\hat{T}_{i,\alpha}(-\ell)\rangle
    &=\frac{1}{\bar{\chi}_{i}^2}\int \dfrac{dk_\parallel} {2\pi}\,H^{\kappa}(k_\parallel)\,H^{\kappa}(-k_\parallel)\,P_\mathrm{mm}\left(\sqrt{k_\parallel^2+ \left(\frac{\ell+1/2}{\bar{\chi}_{i}}\right)^2}, z_i\right)\nonumber\\
    &\int \dfrac{d^3\boldsymbol{K}} {(2\pi)^3}\,H^{v}(K_\parallel)\,H^{v}(-K_\parallel)
    \,P_{v_\mathrm{r}v_\mathrm{r}}\left(|\boldsymbol{K}|, z_\alpha\right)\,.
\end{align}
The kSZ template autocorrelation is a product of two terms. The first term is the cosmic shear power spectrum in the convergence redshift bin, and the second term is the variance of the radial velocity field in the velocity redshift bin. This greatly simplifies the addition of shape-noise term (Eq.~\eqref{eq:shapenoise}) and velocity reconstruction noise (Eq.~\eqref{eq:vnoise}), which can be added independently to the cosmic shear angular power spectrum and the velocity power spectrum to obtain the covariance matrix. 

\section{Toy model for the optimum kernel width}
\label{app:toy}

We build a minimal model that captures the two competing effects that set the optimum nulled-kernel width in Fig.~\ref{fig:snr_vs_width}: the decorrelation
of radial velocities between bins, which favors narrow kernels, and the suppression of the nulled-kernel amplitude with decreasing width, which favors wide kernels. App.~\ref{app:toy-crosscov} adds a note on the
correlated shape noise between nulled bins, which is not captured by the same-bin model below but drives the additional gain shown in Fig.~\ref{fig:snr_crossbins_vs_width}.

\subsection{Setup}
\label{app:toy-setup}

Consider a comoving width $2\Delta\chi$ split into two adjacent sub-bins of width $\Delta\chi$ each. Let $\delta_1, \delta_2$ denote the projected density contrasts and $v^r_1, v^r_2$ the radial velocities in the two sub-bins. We suppress all projection kernels except for an overall kernel height. For simplicity, we assume the densities and velocities in different bins are uncorrelated on the scales of interest, provided $\Delta\chi$ exceeds the velocity correlation length
$\sim 100\,\mathrm{Mpc}$:
\begin{equation}
\langle\delta_1\delta_2\rangle = \langle v^r_1 v^r_2\rangle = 0\,,
\qquad
\langle\delta_i\delta_i\rangle = 2\langle\delta\delta\rangle\,,
\qquad
\langle v^r_i v^r_i\rangle = 2\langle v^r v^r\rangle\,,
\end{equation}
where $\delta = \tfrac12(\delta_1+\delta_2)$ and
$v^r = \tfrac12(v^r_1+v^r_2)$ are the wide-bin fields. Each sub-bin density carries shot noise $2/\bar n$; the wide bin carries $1/\bar n$. The true signal, taking a
constant kSZ kernel, is given by:
\begin{equation}
T = \delta_1 v^r_1 + \delta_2 v^r_2\,.
\end{equation}
Finally, we model the nulling operation by a kernel height $\propto \Delta\chi$: the nulling cancellation suppresses the \emph{signal} amplitude of a nulled kernel in proportion to its width, whereas the shape noise, which adds in quadrature with $\mathcal{O}(1)$ coefficients ($p_{ii}=1$ in Eq.~\eqref{eq:nulling1}), is not suppressed. A nulled sub-bin template density is therefore $\tilde\delta_i = \tfrac12\,\delta_i + n_i$ with
$\langle n_i^2\rangle = 2/\bar n$ unsuppressed.

\subsection{Optimum bin width}
\label{app:toy-widths}

\paragraph{One wide bin:} The template is
$\hat T_a = \delta\, v^r = \tfrac14(\delta_1+\delta_2)(v^r_1+v^r_2)$. Using
the correlators above,
\begin{equation}
\langle T\hat T_a\rangle = 2\langle\delta\delta\rangle\langle v^rv^r\rangle\,,
\qquad
\langle \hat T_a\hat T_a\rangle
 = \langle v^rv^r\rangle\!\left(\langle\delta\delta\rangle
 + \frac{1}{\bar n}\right).
\end{equation}
With the Gaussian covariance, the per-mode significance is
\begin{equation}
\mathrm{SNR}^2_a
 = \frac{\langle T\hat T_a\rangle^2}
        {\langle\hat T_a\hat T_a\rangle\langle TT\rangle
         + \langle T\hat T_a\rangle^2}\,.
\label{eq:toy-snra}
\end{equation}

\paragraph{Two narrow bins with nulling:} 
The template vector is $\hat T_{b,i} = \tilde\delta_i\, v^r_i$, the signal is halved by the kernel
height, $\langle T\hat T_{b,i}\rangle = 2\langle\delta\delta\rangle\langle
v^rv^r\rangle$, while the template variance becomes
$\langle \hat T_{b,i}\hat T_{b,i}\rangle = \langle v^rv^r\rangle (\langle\delta\delta\rangle + 4/\bar n)$: the cosmic-variance part is suppressed by $(\Delta\chi)^2$ but the shape-noise part is not. The SNR in the two regimes becomes,
\begin{align}
\text{shape noise dominated:}\quad
&\mathrm{SNR}^2_b = \tfrac12\,\mathrm{SNR}^2_a
&&\Rightarrow\quad \mathrm{SNR} \propto \sqrt{\Delta\chi}\,,
\label{eq:toy-shot}\\
\text{cosmic-variance dominated:}\quad
&\mathrm{SNR}^2_b = 2\,\mathrm{SNR}^2_a
&&\Rightarrow\quad \mathrm{SNR} \propto \frac{1}{\sqrt{\Delta\chi}}\,,
\label{eq:toy-cv}
\end{align}
When cosmic variance dominates, narrowing the
kernels wins, and when shape noise takes over, the wider kernels become optimum. 

\subsection{Cross-bin covariance}
\label{app:toy-crosscov}

The model above retains only same-bin templates with independent noise. In the full analysis the nulled kernels share source bins (Eq.~\eqref{eq:shapenoise}), so the shape noise of adjacent nulled bins is correlated, and cross-bin templates $(\mathcal{K}_i,\mathcal{V}_\alpha)$
with non-overlapping kernels carry this correlated noise while contributing negligible signal (see Fig.~\ref{fig:snr_crossbins_vs_width}). The minimal example is a two-template data vector with signal $\boldsymbol{a} = (s, 0)$ and noise correlation $\rho$:
\begin{equation}
\mathbf{C} = \sigma^2
\begin{pmatrix} 1 & \rho \\ \rho & 1 \end{pmatrix}
\quad\Longrightarrow\quad
\mathrm{SNR}^2_{\rm all}
 = \boldsymbol{a}^\top \mathbf{C}^{-1} \boldsymbol{a}
 = \frac{s^2}{\sigma^2\,(1-\rho^2)}
 = \frac{\mathrm{SNR}^2_{\rm same}}{1-\rho^2}\,.
\label{eq:toy-noisemonitor}
\end{equation}
The gain depends on $\rho^2$, so the negative correlations induced by the nulling coefficients (the blue secondary diagonals in Fig.~\ref{fig:snr_crossbins_vs_width}) are as beneficial as positive ones. Narrower kernels produce more nulled bins and hence more noise-correlated partners, so this gain compounds toward small $\Delta\chi$, consistent with the width dependence of the all-bins to same-bin SNR ratio in Fig.~\ref{fig:snr_crossbins_vs_width}.

\section{Reconstructed velocity correlation coefficient}
\label{app:recon_vel}
\begin{figure}[t]
    \centering
    \includegraphics[width=\linewidth]{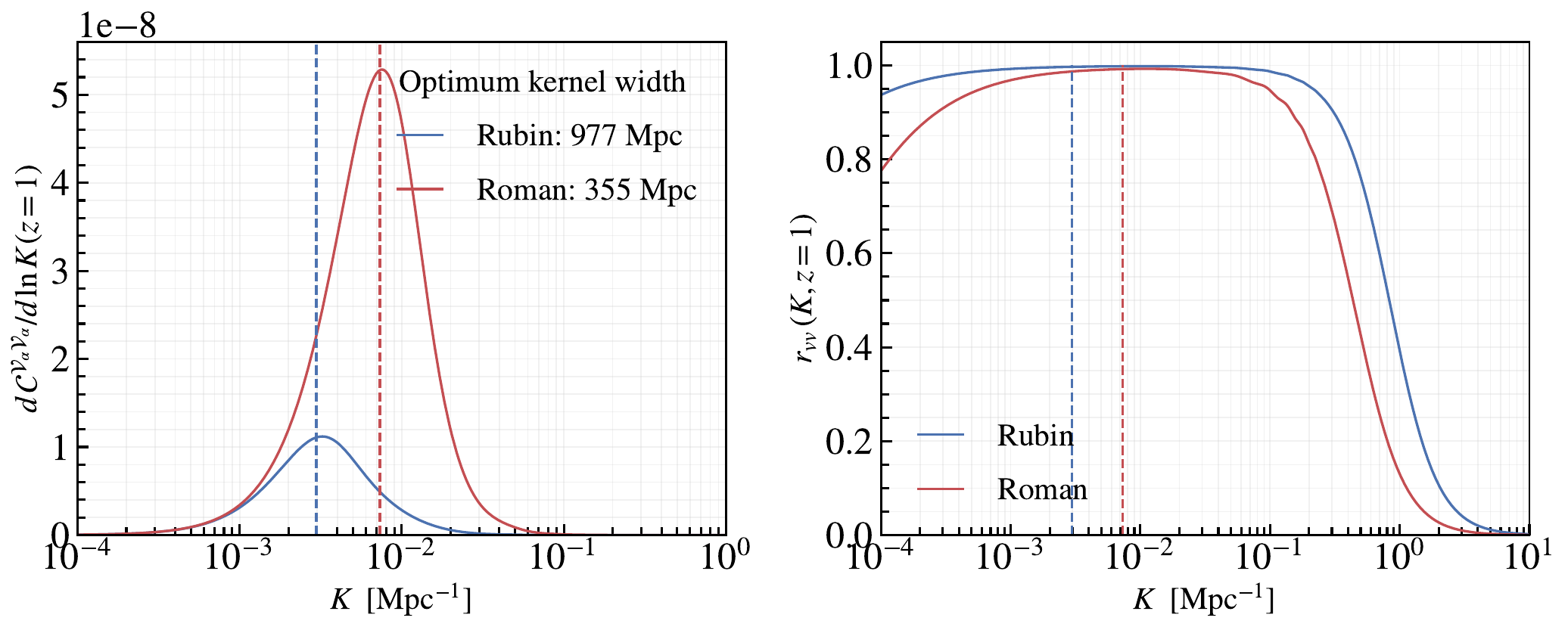}
    \caption{\emph{Left:} Differential contribution to the velocity variance per logarithmic wavenumber, $d\mathcal{C}^{\mathcal{V}_\alpha\mathcal{V}_\alpha}/d\ln K$, evaluated at $z = 1$ for the optimum kernel widths of the two optimum survey configurations (977 Mpc for Rubin Y10, 355 Mpc for Roman KL). Vertical dashed lines mark the $K$ mode corresponding to the peak of the velocity variance dominated by modes comparable to the kernel width, which corresponds to the wavenumbers at which the velocity-reconstruction noise matters the most. \emph{Right:} Velocity-reconstruction correlation coefficient at $z = 1$. Both surveys achieve $r_{vv} \simeq 1$ at the wavenumbers that dominate the velocity variance (dashed lines).}
    \label{fig:vel_noise_mode}
\end{figure}
The correlation coefficient between true and reconstructed velocities in the presence of photo-$z$ error is given by:
\begin{align}
    r_{vv}(k,\mu, z) 
    &= \sqrt{\frac{P_\mathrm{mm}(k, z)S^2(k\mu, \sigma_\chi)}{P_\mathrm{mm}(k, z)S^2(k\mu, \sigma_\chi)+1/\left(\bar{n}_g(z)\,b(z)^2\right)}}\,
    \label{eq:corrcoeff_photoz}
\end{align}
where $S(k\mu) = \exp(-(k\mu\sigma_\chi)^2/2)$ and $\sigma_\chi = (c/H(z))\,\sigma_z$. In Fig.~\ref{fig:vel_noise_mode}, we show that velocity reconstruction is perfect and $r_{vv}$ is close to unity on scales corresponding to the optimum SNR configuration at $z=1$.
\section{Signal and Covariance matrices for optimum configuration}
\label{app:snr_plots}
\begin{figure}[t]
    \centering
    \includegraphics[width=\linewidth]{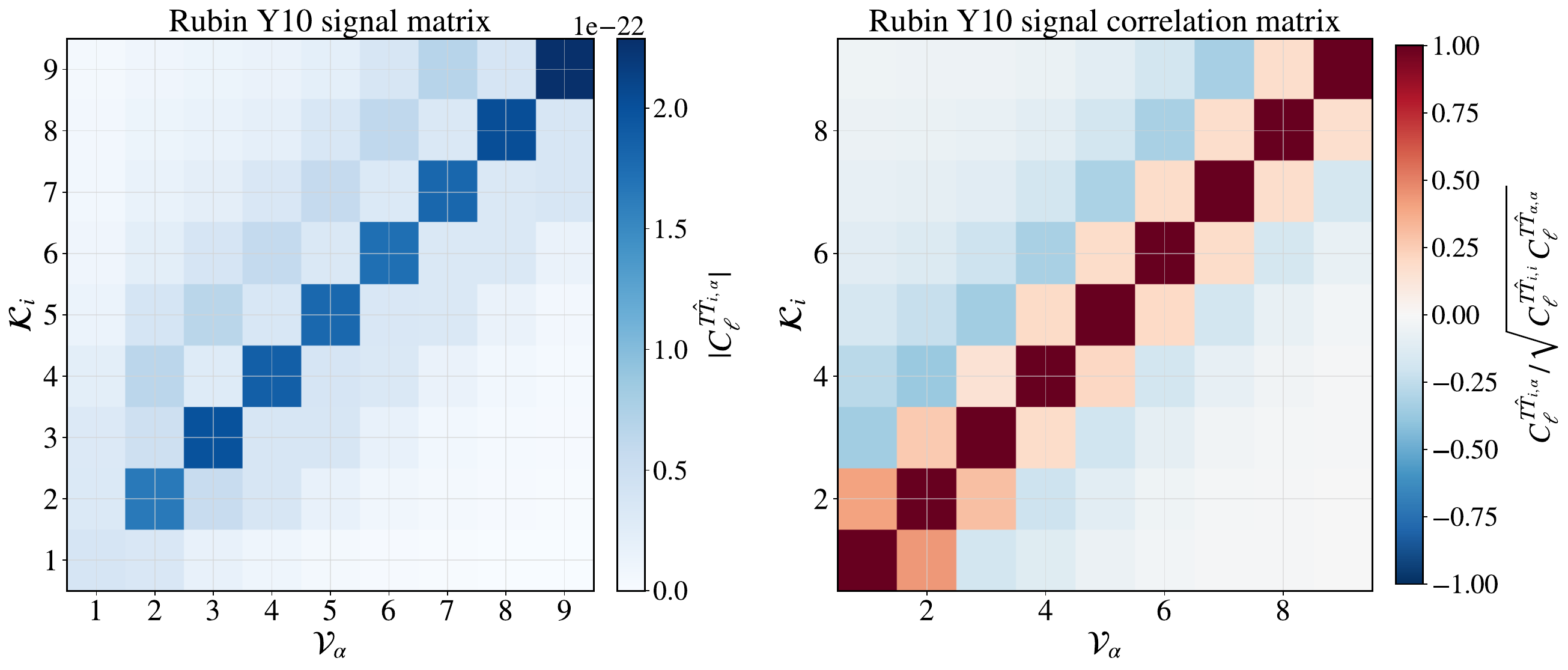}\\[2pt]
    \includegraphics[width=\linewidth]{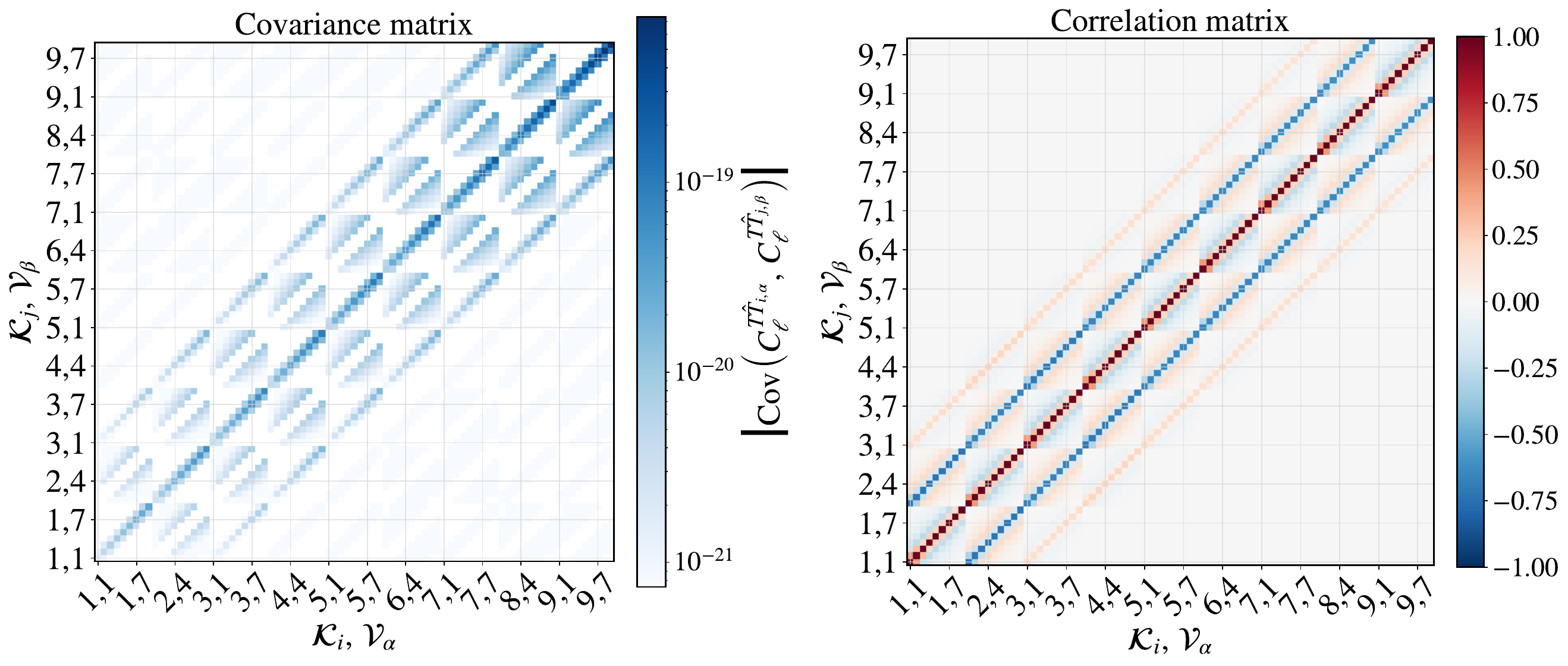}
    \caption{\emph{Top:} Absolute value of the cross-correlation signal and signal correlation matrix. \emph{Bottom:} Absolute value of the covariance matrix and corresponding correlation matrix computed at $\ell=3000$ for optimum survey configuration: 977 Mpc wide kernels for Rubin Y10. The corresponding nulled kernels are shown in the top panel of Fig.~\ref{fig:nulling}.}
    \label{fig:rubin_signal_corr_mat}
\end{figure}

\begin{figure}[t]
    \centering
    \includegraphics[width=\linewidth]{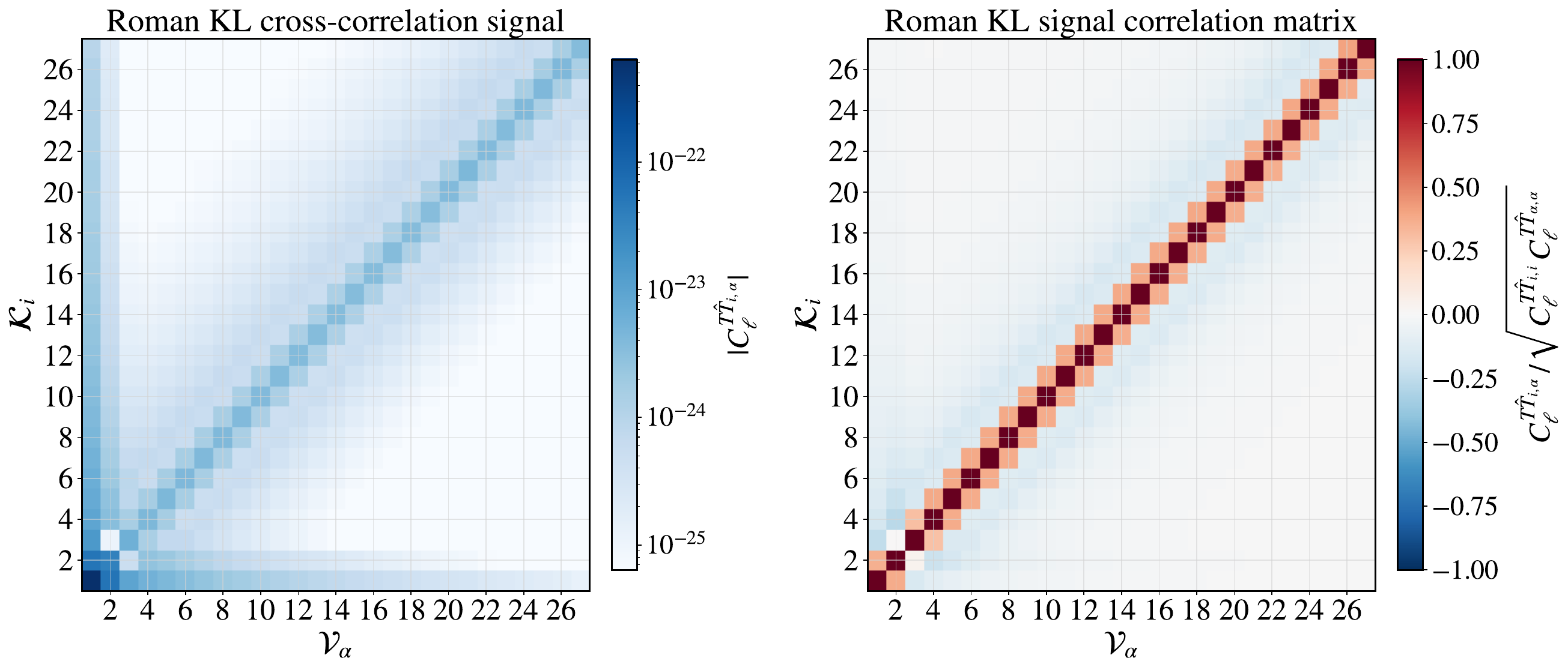}
    \includegraphics[width=\linewidth]{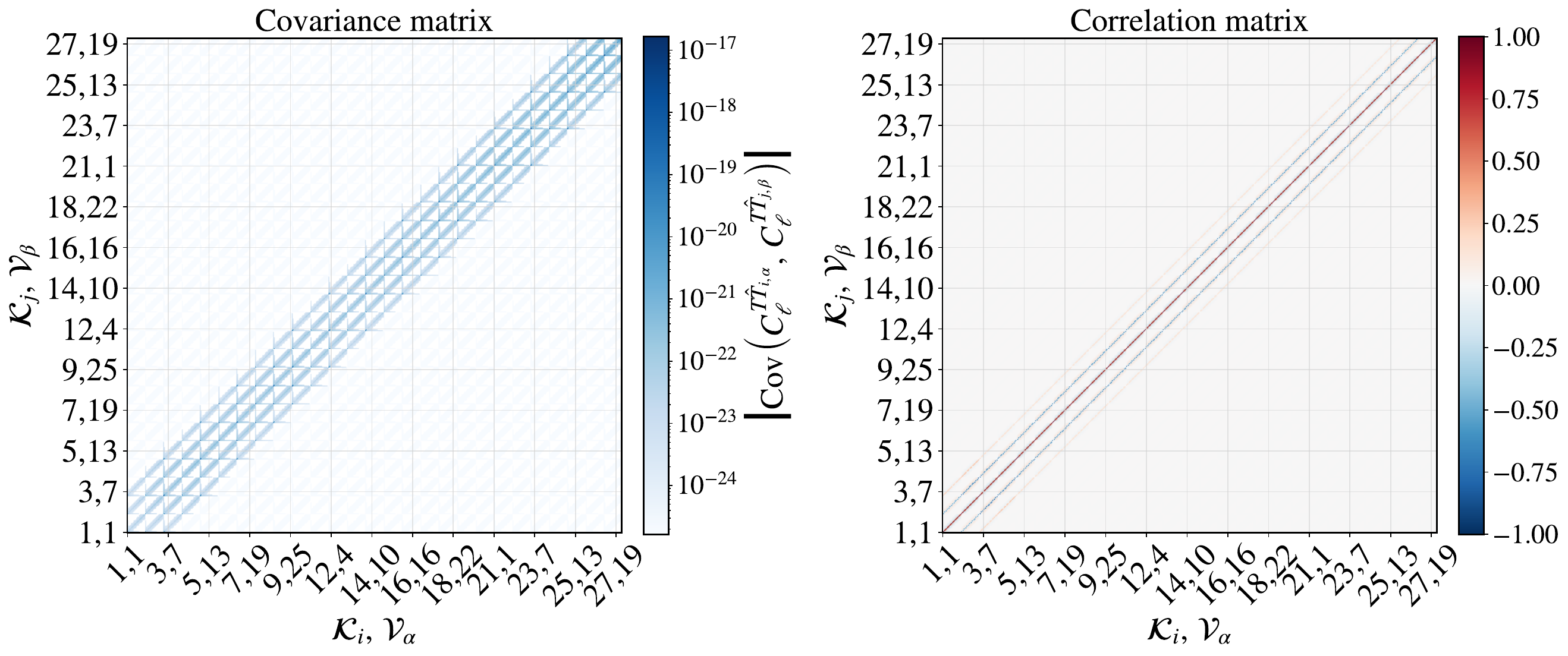}
    \caption{\emph{Top:} Absolute value of the cross-correlation signal and signal correlation matrix, \emph{Bottom:} Absolute value of the covariance matrix and corresponding correlation matrix computed at $\ell=3000$ for optimum survey configuration: 355 Mpc wide kernels for Roman KL. The corresponding nulled kernels are shown in the bottom panel of Fig.~\ref{fig:nulling}.}
    \label{fig:roman_signal_corr_mat}
\end{figure}

\newpage

\bibliography{reference.bib}

@ARTICLE{LSST-DESC-SRD_2018,
       author = {{The LSST Dark Energy Science Collaboration} and {Mandelbaum}, Rachel and {Eifler et al.}, Tim},
        title = "{The LSST Dark Energy Science Collaboration (DESC) Science Requirements Document}",
      journal = {arXiv e-prints},
         year = 2018,
        month = sep,
          eid = {arXiv:1809.01669},
        pages = {arXiv:1809.01669},
          doi = {10.48550/arXiv.1809.01669},
archivePrefix = {arXiv},
       eprint = {1809.01669},
 primaryClass = {astro-ph.CO},
       adsurl = {https://ui.adsabs.harvard.edu/abs/2018arXiv180901669T}
}

@ARTICLE{Zhai_Wang2021,
       author = {{Zhai}, Zhongxu and {Wang}, Yun and {Benson}, Andrew and {Chuang}, Chia-Hsun and {Yepes}, Gustavo},
        title = "{Linear bias and halo occupation distribution of emission-line galaxies from Nancy Grace Roman Space Telescope}",
      journal = {\mnras},
         year = 2021,
        month = aug,
       volume = {505},
       number = {2},
        pages = {2784-2800},
          doi = {10.1093/mnras/stab1539},
archivePrefix = {arXiv},
       eprint = {2103.11063},
 primaryClass = {astro-ph.CO},
       adsurl = {https://ui.adsabs.harvard.edu/abs/2021MNRAS.505.2784Z}
}

@ARTICLE{Battaglia_2016,
       author = {{Battaglia}, N.},
        title = "{The tau of galaxy clusters}",
      journal = {\jcap},
         year = 2016,
        month = aug,
       volume = {2016},
       number = {8},
          eid = {058},
        pages = {058},
          doi = {10.1088/1475-7516/2016/08/058},
archivePrefix = {arXiv},
       eprint = {1607.02442},
 primaryClass = {astro-ph.CO},
       adsurl = {https://ui.adsabs.harvard.edu/abs/2016JCAP...08..058B}
}

@ARTICLE{Bernardeau_2014,
       author = {{Bernardeau}, Francis and {Nishimichi}, Takahiro and {Taruya}, Atsushi},
        title = "{Cosmic shear full nulling: sorting out dynamics, geometry and systematics}",
      journal = {\mnras},
         year = 2014,
        month = dec,
       volume = {445},
       number = {2},
        pages = {1526-1537},
          doi = {10.1093/mnras/stu1861},
archivePrefix = {arXiv},
       eprint = {1312.0430},
 primaryClass = {astro-ph.CO},
       adsurl = {https://ui.adsabs.harvard.edu/abs/2014MNRAS.445.1526B}
}

@ARTICLE{HutererWhite_2005,
       author = {{Huterer}, Dragan and {White}, Martin},
        title = "{Nulling tomography with weak gravitational lensing}",
      journal = {\prd},
         year = 2005,
        month = aug,
       volume = {72},
       number = {4},
          eid = {043002},
        pages = {043002},
          doi = {10.1103/PhysRevD.72.043002},
archivePrefix = {arXiv},
       eprint = {astro-ph/0501451},
 primaryClass = {astro-ph},
       adsurl = {https://ui.adsabs.harvard.edu/abs/2005PhRvD..72d3002H}
}

@ARTICLE{XuEifler2023,
       author = {{Xu}, Jiachuan and {Eifler}, Tim and {Huff}, Eric and {Pranjal}, R.~S. and {Huang}, Hung-Jin and {Everett}, Spencer and {Krause}, Elisabeth},
        title = "{Kinematic lensing with the Roman Space Telescope}",
      journal = {\mnras},
         year = 2023,
        month = feb,
       volume = {519},
       number = {2},
        pages = {2535-2551},
          doi = {10.1093/mnras/stac3685},
archivePrefix = {arXiv},
       eprint = {2201.00739},
 primaryClass = {astro-ph.CO},
       adsurl = {https://ui.adsabs.harvard.edu/abs/2023MNRAS.519.2535X}
}

@ARTICLE{vanDaalen2011,
       author = {{van Daalen}, Marcel P. and {Schaye}, Joop and {Booth}, C.~M. and {Dalla Vecchia}, Claudio},
        title = "{The effects of galaxy formation on the matter power spectrum: a challenge for precision cosmology}",
      journal = {\mnras},
         year = 2011,
        month = aug,
       volume = {415},
       number = {4},
        pages = {3649-3665},
          doi = {10.1111/j.1365-2966.2011.18981.x},
archivePrefix = {arXiv},
       eprint = {1104.1174},
 primaryClass = {astro-ph.CO},
       adsurl = {https://ui.adsabs.harvard.edu/abs/2011MNRAS.415.3649V}
}

@ARTICLE{Chisari2019,
       author = {{Chisari}, Nora Elisa and {Mead}, Alexander J. and {Joudaki}, Shahab and {Ferreira}, Pedro G. and {Schneider}, Aurel and {Mohr}, Joseph and {Tr{\"o}ster}, Tilman and {Alonso}, David and {McCarthy}, Ian G. and {Martin-Alvarez}, Sergio and {Devriendt}, Julien and {Slyz}, Adrianne and {van Daalen}, Marcel P.},
        title = "{Modelling baryonic feedback for survey cosmology}",
      journal = {The Open Journal of Astrophysics},
         year = 2019,
        month = jun,
       volume = {2},
       number = {1},
          eid = {4},
        pages = {4},
          doi = {10.21105/astro.1905.06082},
archivePrefix = {arXiv},
       eprint = {1905.06082},
 primaryClass = {astro-ph.CO},
       adsurl = {https://ui.adsabs.harvard.edu/abs/2019OJAp....2E...4C}
}

@ARTICLE{Huang2019,
       author = {{Huang}, Hung-Jin and {Eifler}, Tim and {Mandelbaum}, Rachel and {Dodelson}, Scott},
        title = "{Modelling baryonic physics in future weak lensing surveys}",
      journal = {\mnras},
         year = 2019,
        month = sep,
       volume = {488},
       number = {2},
        pages = {1652-1678},
          doi = {10.1093/mnras/stz1714},
archivePrefix = {arXiv},
       eprint = {1809.01146},
 primaryClass = {astro-ph.CO},
       adsurl = {https://ui.adsabs.harvard.edu/abs/2019MNRAS.488.1652H}
}

@ARTICLE{Soergel2016,
       author = {{Soergel}, B. and {Flender}, S. and {Story}, K.~T. and {DES Collaboration} and {SPT Collaboration}},
        title = "{Detection of the kinematic Sunyaev-Zel'dovich effect with DES Year 1 and SPT}",
      journal = {\mnras},
         year = 2016,
        month = sep,
       volume = {461},
       number = {3},
        pages = {3172-3193},
          doi = {10.1093/mnras/stw1455},
archivePrefix = {arXiv},
       eprint = {1603.03904},
 primaryClass = {astro-ph.CO},
       adsurl = {https://ui.adsabs.harvard.edu/abs/2016MNRAS.461.3172S}
}

@ARTICLE{Hand2012,
       author = {{Hand}, Nick and {Addison}, Graeme E. and {Aubourg}, Eric and {Battaglia et al.}, Nick},
        title = "{Evidence of Galaxy Cluster Motions with the Kinematic Sunyaev-Zel'dovich Effect}",
      journal = {\prl},
         year = 2012,
        month = jul,
       volume = {109},
       number = {4},
          eid = {041101},
        pages = {041101},
          doi = {10.1103/PhysRevLett.109.041101},
archivePrefix = {arXiv},
       eprint = {1203.4219},
 primaryClass = {astro-ph.CO},
       adsurl = {https://ui.adsabs.harvard.edu/abs/2012PhRvL.109d1101H}
}

@ARTICLE{Schaan2021,
       author = {{Schaan}, Emmanuel and {Ferraro}, Simone and {Amodeo}, Stefania and {Battaglia}, Nicholas  and {Atacama Cosmology Telescope Collaboration}},
        title = "{Atacama Cosmology Telescope: Combined kinematic and thermal Sunyaev-Zel'dovich measurements from BOSS CMASS and LOWZ halos}",
      journal = {\prd},
         year = 2021,
        month = mar,
       volume = {103},
       number = {6},
          eid = {063513},
        pages = {063513},
          doi = {10.1103/PhysRevD.103.063513},
archivePrefix = {arXiv},
       eprint = {2009.05557},
 primaryClass = {astro-ph.CO},
       adsurl = {https://ui.adsabs.harvard.edu/abs/2021PhRvD.103f3513S}
}

@ARTICLE{Calafut2021,
       author = {{Calafut}, V. and {Gallardo}, P.~A. and {Vavagiakis et al.}, E.~M.},
        title = "{The Atacama Cosmology Telescope: Detection of the pairwise kinematic Sunyaev-Zel'dovich effect with SDSS DR15 galaxies}",
      journal = {\prd},
         year = 2021,
        month = aug,
       volume = {104},
       number = {4},
          eid = {043502},
        pages = {043502},
          doi = {10.1103/PhysRevD.104.043502},
archivePrefix = {arXiv},
       eprint = {2101.08374},
 primaryClass = {astro-ph.CO},
       adsurl = {https://ui.adsabs.harvard.edu/abs/2021PhRvD.104d3502C}
}

@ARTICLE{Hadzhiyska2024,
       author = {{Hadzhiyska}, B. and {Ferraro}, S. and {Ried Guachalla}, B. and {Schaan et al.}, E.},
        title = "{Evidence for large baryonic feedback at low and intermediate redshifts from kinematic Sunyaev-Zel'dovich observations with ACT and DESI photometric galaxies}",
      journal = {\prd},
         year = 2025,
        month = oct,
       volume = {112},
       number = {8},
          eid = {083509},
        pages = {083509},
          doi = {10.1103/kclp-x5j1},
archivePrefix = {arXiv},
       eprint = {2407.07152},
 primaryClass = {astro-ph.CO},
       adsurl = {https://ui.adsabs.harvard.edu/abs/2025PhRvD.112h3509H}
}

@ARTICLE{Sunyaev1980,
       author = {{Sunyaev}, R.~A. and {Zeldovich}, Ia. B.},
        title = "{Microwave background radiation as a probe of the contemporary structure and history of the universe}",
      journal = {\araa},
         year = 1980,
        month = jan,
       volume = {18},
        pages = {537-560},
          doi = {10.1146/annurev.aa.18.090180.002541},
       adsurl = {https://ui.adsabs.harvard.edu/abs/1980ARA&A..18..537S}
}

@ARTICLE{Qu2026,
       author = {{Qu}, F.~J. and {Ried Guachalla}, B. and {Schaan}, E. and {Hadzhiyska et al.}, B.},
        title = "{Precision Kinematic Sunyaev--Zel'dovich Measurements Across Halo Mass and Redshift with DESI DR2 and ACT DR6: Part I. Luminous Red Galaxies}",
      journal = {arXiv e-prints},
         year = 2026,
        month = apr,
          eid = {arXiv:2604.19744},
        pages = {arXiv:2604.19744},
          doi = {10.48550/arXiv.2604.19744},
archivePrefix = {arXiv},
       eprint = {2604.19744},
 primaryClass = {astro-ph.CO},
       adsurl = {https://ui.adsabs.harvard.edu/abs/2026arXiv260419744Q}
}

@ARTICLE{Hill2016,
       author = {{Hill}, J. Colin and {Ferraro}, Simone and {Battaglia}, Nick and {Liu}, Jia and {Spergel}, David N.},
        title = "{Kinematic Sunyaev-Zel'dovich Effect with Projected Fields: A Novel Probe of the Baryon Distribution with Planck, WMAP, and WISE Data}",
      journal = {\prl},
         year = 2016,
        month = jul,
       volume = {117},
       number = {5},
          eid = {051301},
        pages = {051301},
          doi = {10.1103/PhysRevLett.117.051301},
archivePrefix = {arXiv},
       eprint = {1603.01608},
 primaryClass = {astro-ph.CO},
       adsurl = {https://ui.adsabs.harvard.edu/abs/2016PhRvL.117e1301H}
}

@ARTICLE{Kusiak2021,
       author = {{Kusiak}, Aleksandra and {Bolliet}, Boris and {Ferraro}, Simone and {Hill}, J. Colin and {Krolewski}, Alex},
        title = "{Constraining the baryon abundance with the kinematic Sunyaev-Zel'dovich effect: Projected-field detection using P l a n c k , W M A P , and u n W I S E}",
      journal = {\prd},
         year = 2021,
        month = aug,
       volume = {104},
       number = {4},
          eid = {043518},
        pages = {043518},
          doi = {10.1103/PhysRevD.104.043518},
archivePrefix = {arXiv},
       eprint = {2102.01068},
 primaryClass = {astro-ph.CO},
       adsurl = {https://ui.adsabs.harvard.edu/abs/2021PhRvD.104d3518K}
}

@ARTICLE{Bolliet2023,
       author = {{Bolliet}, Boris and {Colin Hill}, J. and {Ferraro}, Simone and {Kusiak}, Aleksandra and {Krolewski}, Alex},
        title = "{Projected-field kinetic Sunyaev-Zel'dovich Cross-correlations: halo model and forecasts}",
      journal = {\jcap},
         year = 2023,
        month = mar,
       volume = {2023},
       number = {3},
          eid = {039},
        pages = {039},
          doi = {10.1088/1475-7516/2023/03/039},
archivePrefix = {arXiv},
       eprint = {2208.07847},
 primaryClass = {astro-ph.CO},
       adsurl = {https://ui.adsabs.harvard.edu/abs/2023JCAP...03..039B}
}

@ARTICLE{Smith2018,
       author = {{Smith}, Kendrick M. and {Madhavacheril}, Mathew S. and {M{\"u}nchmeyer}, Moritz and {Ferraro}, Simone and {Giri}, Utkarsh and {Johnson}, Matthew C.},
        title = "{KSZ tomography and the bispectrum}",
      journal = {arXiv e-prints},
         year = 2018,
        month = oct,
          eid = {arXiv:1810.13423},
        pages = {arXiv:1810.13423},
          doi = {10.48550/arXiv.1810.13423},
archivePrefix = {arXiv},
       eprint = {1810.13423},
 primaryClass = {astro-ph.CO},
       adsurl = {https://ui.adsabs.harvard.edu/abs/2018arXiv181013423S}
}

@ARTICLE{Wayland2026,
       author = {{Wayland}, Amy and {Alonso}, David and {Posta}, Adrien La},
        title = "{Detailed theoretical modelling of the kinetic Sunyaev-Zel'dovich stacking power spectrum}",
      journal = {\jcap},
         year = 2026,
        month = jan,
       volume = {2026},
       number = {1},
          eid = {015},
        pages = {015},
          doi = {10.1088/1475-7516/2026/01/015},
archivePrefix = {arXiv},
       eprint = {2509.18732},
 primaryClass = {astro-ph.CO},
       adsurl = {https://ui.adsabs.harvard.edu/abs/2026JCAP...01..015W}
}

@ARTICLE{Planck2020,
       author = {{Planck Collaboration} and {Aghanim}, N. and et al.},
        title = "{Planck 2018 results. VI. Cosmological parameters}",
      journal = {\aap},
         year = 2020,
        month = sep,
       volume = {641},
          eid = {A6},
        pages = {A6},
          doi = {10.1051/0004-6361/201833910},
archivePrefix = {arXiv},
       eprint = {1807.06209},
 primaryClass = {astro-ph.CO},
       adsurl = {https://ui.adsabs.harvard.edu/abs/2020A&A...641A...6P}
}

@misc{guachalla2025backlightingextendedgashalos,
      title={Backlighting extended gas halos around luminous red galaxies: kinematic Sunyaev-Zel'dovich effect from DESI Y1 x ACT}, 
      author={Bernardita Ried Guachalla and Emmanuel Schaan and Boryana Hadzhiyska and Simone Ferraro and Jessica N. Aguilar and Steven Ahlen and Nicholas Battaglia and Davide Bianchi and Richard Bond and David Brooks and Todd Claybaugh and William R. Coulton and Axel de la Macorra and Mark J. Devlin and Arjun Dey and Peter Doel and Jo Dunkley and Kevin Fanning and Jaime Forero-Romero and Enrique Gazta{\~n}aga and Satya Gontcho A Gontcho and Gaston Gutierrez and Julien Guy and J. Colin Hill and Klaus Honscheid and Stephanie Juneau and Theodore Kisner and Anthony Kremin and Andrew Lambert and Martin Landriau and Laurent Le Guillou and Niall MacCrann and Marc Manera and Aaron Meisner and Ramon Miquel and Kavilan Moodley and John Moustakas and Tony Mroczkowski and Adam D. Myers and Michael D. Niemack and Gustavo Niz and Nathalie Palanque-Delabrouille and Will Percival and Ignasi P{\'e}rez-R{\`a}fols and Claire Poppett and Francisco Prada and Frank J. Qu and Graziano Rossi and Eusebio Sanchez and David Schlegel and Michael Schubnell and Hee-Jong Seo and Crist{\'o}bal Sif{\'o}n and David N. Spergel and David Sprayberry and Gregory Tarl{\'e} and Mariana Vargas-Maga{\~n}a and Eve M. Vavagiakis and Benjamin A. Weaver and Edward J. Wollack and Pauline Zarrouk},
      year={2025},
      eprint={2503.19870},
      archivePrefix={arXiv},
      primaryClass={astro-ph.GA},
      url={https://arxiv.org/abs/2503.19870}, 
}

@misc{hadzhiyska2025evidencelargebaryonicfeedback,
      title={Evidence for large baryonic feedback at low and intermediate redshifts from kinematic Sunyaev-Zel'dovich observations with ACT and DESI photometric galaxies}, 
      author={B. Hadzhiyska and S. Ferraro and B. Ried Guachalla and E. Schaan and J. Aguilar and N. Battaglia and J. R. Bond and D. Brooks and E. Calabrese and S. K. Choi and T. Claybaugh and W. R. Coulton and K. Dawson and M. Devlin and B. Dey and P. Doel and A. J. Duivenvoorden and J. Dunkley and G. S. Farren and A. Font-Ribera and J. E. Forero-Romero and P. A. Gallardo and E. Gazta{\~n}aga and S. Gontcho Gontcho and M. Gralla and L. Le Guillou and G. Gutierrez and J. Guy and J. C. Hill and R. Hlo{\v{z}}ek and K. Honscheid and S. Juneau and T. Kisner and A. Kremin and M. Landriau and R. H. Liu and T. Louis and N. MacCrann and A. de Macorra and M. Madhavacheril and M. Manera and A. Meisner and R. Miquel and K. Moodley and J. Moustakas and T. Mroczkowski and S. Naess and J. Newman and M. D. Niemack and G. Niz and L. Page and N. Palanque-Delabrouille and B. Partridge and W. J. Percival and F. Prada and F. J. Qu and G. Rossi and E. Sanchez and D. Schlegel and M. Schubnell and N. Sehgal and H. Seo and C. Sif{\'o}n and D. Spergel and D. Sprayberry and S. Staggs and G. Tarl{\'e} and C. Vargas and E. M. Vavagiakis and B. A. Weaver and E. J. Wollack and R. Zhou and H. Zou},
      year={2025},
      eprint={2407.07152},
      archivePrefix={arXiv},
      primaryClass={astro-ph.CO},
      url={https://arxiv.org/abs/2407.07152}, 
}

@article{PhysRevD.109.103534,
  title = {Velocity reconstruction in the era of DESI and Rubin/LSST. II. Realistic samples on the light cone},
  author = {Hadzhiyska, Boryana and Ferraro, Simone and Ried Guachalla, Bernardita and Schaan, Emmanuel},
  journal = {Phys. Rev. D},
  volume = {109},
  issue = {10},
  pages = {103534},
  numpages = {17},
  year = {2024},
  month = {May},
  publisher = {American Physical Society},
  doi = {10.1103/PhysRevD.109.103534},
  url = {https://link.aps.org/doi/10.1103/PhysRevD.109.103534}
}

@article{Bolliet_2023,
   title={Projected-field kinetic Sunyaev-Zeldovich Cross-correlations: halo model and forecasts},
   volume={2023},
   ISSN={1475-7516},
   url={http://dx.doi.org/10.1088/1475-7516/2023/03/039},
   DOI={10.1088/1475-7516/2023/03/039},
   number={03},
   journal={Journal of Cosmology and Astroparticle Physics},
   publisher={IOP Publishing},
   author={Bolliet, Boris and Colin Hill, J. and Ferraro, Simone and Kusiak, Aleksandra and Krolewski, Alex},
   year={2023},
   month=mar, pages={039} }

@article{ACT:2023kun,
    author = "Madhavacheril, Mathew S. and others",
    collaboration = "ACT",
    title = "{The Atacama Cosmology Telescope: DR6 Gravitational Lensing Map and Cosmological Parameters}",
    eprint = "2304.05203",
    archivePrefix = "arXiv",
    primaryClass = "astro-ph.CO",
    reportNumber = "FERMILAB-PUB-23-206-PPD",
    doi = "10.3847/1538-4357/acff5f",
    journal = "Astrophys. J.",
    volume = "962",
    number = "2",
    pages = "113",
    year = "2024"
}

@article{PhysRevD.109.103533,
  title = {Velocity reconstruction in the era of DESI and Rubin/LSST. I. Exploring spectroscopic, photometric, and hybrid samples},
  author = {Ried Guachalla, Bernardita and Schaan, Emmanuel and Hadzhiyska, Boryana and Ferraro, Simone},
  journal = {Phys. Rev. D},
  volume = {109},
  issue = {10},
  pages = {103533},
  numpages = {20},
  year = {2024},
  month = {May},
  publisher = {American Physical Society},
  doi = {10.1103/PhysRevD.109.103533},
  url = {https://link.aps.org/doi/10.1103/PhysRevD.109.103533}
}

@article{PhysRevLett.109.041101,
  title = {Evidence of Galaxy Cluster Motions with the Kinematic Sunyaev-Zel'dovich Effect},
  author = {Hand, Nick and Addison, Graeme E. and Aubourg, Eric and Battaglia, Nick and Battistelli, Elia S. and Bizyaev, Dmitry and Bond, J. Richard and Brewington, Howard and Brinkmann, Jon and Brown, Benjamin R. and Das, Sudeep and Dawson, Kyle S. and Devlin, Mark J. and Dunkley, Joanna and Dunner, Rolando and Eisenstein, Daniel J. and Fowler, Joseph W. and Gralla, Megan B. and Hajian, Amir and Halpern, Mark and Hilton, Matt and Hincks, Adam D. and Hlozek, Ren\'ee and Hughes, John P. and Infante, Leopoldo and Irwin, Kent D. and Kosowsky, Arthur and Lin, Yen-Ting and Malanushenko, Elena and Malanushenko, Viktor and Marriage, Tobias A. and Marsden, Danica and Menanteau, Felipe and Moodley, Kavilan and Niemack, Michael D. and Nolta, Michael R. and Oravetz, Daniel and Page, Lyman A. and Palanque-Delabrouille, Nathalie and Pan, Kaike and Reese, Erik D. and Schlegel, David J. and Schneider, Donald P. and Sehgal, Neelima and Shelden, Alaina and Sievers, Jon and Sif\'on, Crist\'obal and Simmons, Audrey and Snedden, Stephanie and Spergel, David N. and Staggs, Suzanne T. and Swetz, Daniel S. and Switzer, Eric R. and Trac, Hy and Weaver, Benjamin A. and Wollack, Edward J. and Yeche, Christophe and Zunckel, Caroline},
  journal = {Phys. Rev. Lett.},
  volume = {109},
  issue = {4},
  pages = {041101},
  numpages = {6},
  year = {2012},
  month = {Jul},
  publisher = {American Physical Society},
  doi = {10.1103/PhysRevLett.109.041101},
  url = {https://link.aps.org/doi/10.1103/PhysRevLett.109.041101}
}

@article{PhysRevLett.115.191301,
  title = {Evidence of the Missing Baryons from the Kinematic Sunyaev-Zeldovich Effect in Planck Data},
  author = {Hern\'andez-Monteagudo, Carlos and Ma, Yin-Zhe and Kitaura, Francisco S. and Wang, Wenting and G\'enova-Santos, Ricardo and Mac\'{\i}as-P\'erez, Juan and Herranz, Diego},
  journal = {Phys. Rev. Lett.},
  volume = {115},
  issue = {19},
  pages = {191301},
  numpages = {5},
  year = {2015},
  month = {Nov},
  publisher = {American Physical Society},
  doi = {10.1103/PhysRevLett.115.191301},
  url = {https://link.aps.org/doi/10.1103/PhysRevLett.115.191301}
}

@article{Bernardis_2017,
   title={Detection of the pairwise kinematic Sunyaev-Zeldovich effect with BOSS DR11 and the Atacama Cosmology Telescope},
   volume={2017},
   ISSN={1475-7516},
   url={http://dx.doi.org/10.1088/1475-7516/2017/03/008},
   DOI={10.1088/1475-7516/2017/03/008},
   number={03},
   journal={Journal of Cosmology and Astroparticle Physics},
   publisher={IOP Publishing},
   author={Bernardis, F. De and Aiola, S. and Vavagiakis, E.M. and Battaglia, N. and Niemack, M.D. and Beall, J. and Becker, D.T. and Bond, J.R. and Calabrese, E. and Cho, H. and Coughlin, K. and Datta, R. and Devlin, M. and Dunkley, J. and Dunner, R. and Ferraro, S. and Fox, A. and Gallardo, P.A. and Halpern, M. and Hand, N. and Hasselfield, M. and Henderson, S.W. and Hill, J.C. and Hilton, G.C. and Hilton, M. and Hincks, A.D. and Hlozek, R. and Hubmayr, J. and Huffenberger, K. and Hughes, J.P. and Irwin, K.D. and Koopman, B.J. and Kosowsky, A. and Li, D. and Louis, T. and Lungu, M. and Madhavacheril, M.S. and Maurin, L. and McMahon, J. and Moodley, K. and Naess, S. and Nati, F. and Newburgh, L. and Nibarger, J.P. and Page, L.A. and Partridge, B. and Schaan, E. and Schmitt, B. L. and Sehgal, N. and Sievers, J. and Simon, S.M. and Spergel, D.N. and Staggs, S.T. and Stevens, J.R. and Thornton, R.J. and Engelen, A. van and Lanen, J. Van and Wollack, E.J.},
   year={2017},
   month=mar, pages={008--008} }

@article{Soergel_2016,
   title={Detection of the kinematic Sunyaev--Zeldovich effect with DES Year 1 and SPT},
   volume={461},
   ISSN={1365-2966},
   url={http://dx.doi.org/10.1093/mnras/stw1455},
   DOI={10.1093/mnras/stw1455},
   number={3},
   journal={Monthly Notices of the Royal Astronomical Society},
   publisher={Oxford University Press (OUP)},
   author={Soergel, B. and Flender, S. and Story, K. T. and Bleem, L. and Giannantonio, T. and Efstathiou, G. and Rykoff, E. and Benson, B. A. and Crawford, T. and Dodelson, S. and Habib, S. and Heitmann, K. and Holder, G. and Jain, B. and Rozo, E. and Saro, A. and Weller, J. and Abdalla, F. B. and Allam, S. and Annis, J. and Armstrong, R. and Benoit-L{\'e}vy, A. and Bernstein, G. M. and Carlstrom, J. E. and Carnero Rosell, A. and Carrasco Kind, M. and Castander, F. J. and Chiu, I. and Chown, R. and Crocce, M. and Cunha, C. E. and DAndrea, C. B. and da Costa, L. N. and de Haan, T. and Desai, S. and Diehl, H. T. and Dietrich, J. P. and Doel, P. and Estrada, J. and Evrard, A. E. and Flaugher, B. and Fosalba, P. and Frieman, J. and Gaztanaga, E. and Gruen, D. and Gruendl, R. A. and Holzapfel, W. L. and Honscheid, K. and James, D. J. and Keisler, R. and Kuehn, K. and Kuropatkin, N. and Lahav, O. and Lima, M. and Marshall, J. L. and McDonald, M. and Melchior, P. and Miller, C. J. and Miquel, R. and Nord, B. and Ogando, R. and Omori, Y. and Plazas, A. A. and Rapetti, D. and Reichardt, C. L. and Romer, A. K. and Roodman, A. and Saliwanchik, B. R. and Sanchez, E. and Schubnell, M. and Sevilla-Noarbe, I. and Sheldon, E. and Smith, R. C. and Soares-Santos, M. and Sobreira, F. and Stark, A. and Suchyta, E. and Swanson, M. E. C. and Tarle, G. and Thomas, D. and Vieira, J. D. and Walker, A. R. and Whitehorn, N.},
   year={2016},
   month=jun, pages={3172--3193} }

@article{Sugiyama_2018,
   title={A direct measure of free electron gas via the kinematic Sunyaev--Zeldovich effect in Fourier-space analysis},
   volume={475},
   ISSN={1365-2966},
   url={http://dx.doi.org/10.1093/mnras/stx3362},
   DOI={10.1093/mnras/stx3362},
   number={3},
   journal={Monthly Notices of the Royal Astronomical Society},
   publisher={Oxford University Press (OUP)},
   author={Sugiyama, Naonori S and Okumura, Teppei and Spergel, David N},
   year={2018},
   month=jan, pages={3764--3785} }

@article{PhysRevD.104.043502,
  title = {The Atacama Cosmology Telescope: Detection of the pairwise kinematic Sunyaev-Zel'dovich effect with SDSS DR15 galaxies},
  author = {Calafut, V. and Gallardo, P. A. and Vavagiakis, E. M. and Amodeo, S. and Aiola, S. and Austermann, J. E. and Battaglia, N. and Battistelli, E. S. and Beall, J. A. and Bean, R. and Bond, J. R. and Calabrese, E. and Choi, S. K. and Cothard, N. F. and Devlin, M. J. and Duell, C. J. and Duff, S. M. and Duivenvoorden, A. J. and Dunkley, J. and Dunner, R. and Ferraro, S. and Guan, Y. and Hill, J. C. and Hilton, G. C. and Hilton, M. and Hlo\ifmmode \check{z}\else \v{z}\fi{}ek, R. and Huber, Z. B. and Hubmayr, J. and Huffenberger, K. M. and Hughes, J. P. and Koopman, B. J. and Kosowsky, A. and Li, Y. and Lokken, M. and Madhavacheril, M. and McMahon, J. and Moodley, K. and Naess, S. and Nati, F. and Newburgh, L. B. and Niemack, M. D. and Page, L. A. and Partridge, B. and Schaan, E. and Schillaci, A. and Sif\'on, C. and Spergel, D. N. and Staggs, S. T. and Ullom, J. N. and Vale, L. R. and Van Engelen, A. and Van Lanen, J. and Wollack, E. J. and Xu, Z.},
  journal = {Phys. Rev. D},
  volume = {104},
  issue = {4},
  pages = {043502},
  numpages = {16},
  year = {2021},
  month = {Aug},
  publisher = {American Physical Society},
  doi = {10.1103/PhysRevD.104.043502},
  url = {https://link.aps.org/doi/10.1103/PhysRevD.104.043502}
}

@misc{li2024detectionpairwisekineticsunyaevzeldovich,
      title={Detection of pairwise kinetic Sunyaev-Zel'dovich effect with DESI galaxy groups and Planck in Fourier space}, 
      author={Shaohong Li and Yi Zheng and Ziyang Chen and Haojie Xu and Xiaohu Yang},
      year={2024},
      eprint={2401.03507},
      archivePrefix={arXiv},
      primaryClass={astro-ph.CO},
      url={https://arxiv.org/abs/2401.03507}, 
}

@misc{Hadzhiyska:2025egz,
    author = {Hadzhiyska, B. and others},
    title = {{Probing cosmic velocities with the pairwise kinematic Sunyaev-Zel'dovich signal in DESI Bright Galaxy Sample DR1 and ACT DR6}},
    eprint = {2510.14135},
    archivePrefix = {arXiv},
    primaryClass = {astro-ph.CO},
    reportNumber = {FERMILAB-PUB-25-0758-PPD},
    year = {2025}
}

@article{PhysRevD.103.063513,
  title = {Atacama Cosmology Telescope: Combined kinematic and thermal Sunyaev-Zel'dovich measurements from BOSS CMASS and LOWZ halos},
  author = {Schaan, Emmanuel and Ferraro, Simone and Amodeo, Stefania and Battaglia, Nicholas and Aiola, Simone and Austermann, Jason E. and Beall, James A. and Bean, Rachel and Becker, Daniel T. and Bond, Richard J. and Calabrese, Erminia and Calafut, Victoria and Choi, Steve K. and Denison, Edward V. and Devlin, Mark J. and Duff, Shannon M. and Duivenvoorden, Adriaan J. and Dunkley, Jo and D\"unner, Rolando and Gallardo, Patricio A. and Guan, Yilun and Han, Dongwon and Hill, J. Colin and Hilton, Gene C. and Hilton, Matt and Hlo\ifmmode \check{z}\else \v{z}\fi{}ek, Ren\'ee and Hubmayr, Johannes and Huffenberger, Kevin M. and Hughes, John P. and Koopman, Brian J. and MacInnis, Amanda and McMahon, Jeff and Madhavacheril, Mathew S. and Moodley, Kavilan and Mroczkowski, Tony and Naess, Sigurd and Nati, Federico and Newburgh, Laura B. and Niemack, Michael D. and Page, Lyman A. and Partridge, Bruce and Salatino, Maria and Sehgal, Neelima and Schillaci, Alessandro and Sif\'on, Crist\'obal and Smith, Kendrick M. and Spergel, David N. and Staggs, Suzanne and Storer, Emilie R. and Trac, Hy and Ullom, Joel N. and Van Lanen, Jeff and Vale, Leila R. and van Engelen, Alexander and Maga\~na, Mariana Vargas and Vavagiakis, Eve M. and Wollack, Edward J. and Xu, Zhilei},
  collaboration = {Atacama Cosmology Telescope Collaboration},
  journal = {Phys. Rev. D},
  volume = {103},
  issue = {6},
  pages = {063513},
  numpages = {26},
  year = {2021},
  month = {Mar},
  publisher = {American Physical Society},
  doi = {10.1103/PhysRevD.103.063513},
  url = {https://link.aps.org/doi/10.1103/PhysRevD.103.063513}
}

@article{PhysRevD.93.082002,
  title = {Evidence for the kinematic Sunyaev-Zel'dovich effect with the Atacama Cosmology Telescope and velocity reconstruction from the Baryon Oscillation Spectroscopic Survey},
  author = {Schaan, Emmanuel and Ferraro, Simone and Vargas-Maga\~na, Mariana and Smith, Kendrick M. and Ho, Shirley and Aiola, Simone and Battaglia, Nicholas and Bond, J. Richard and De Bernardis, Francesco and Calabrese, Erminia and Cho, Hsiao-Mei and Devlin, Mark J. and Dunkley, Joanna and Gallardo, Patricio A. and Hasselfield, Matthew and Henderson, Shawn and Hill, J. Colin and Hincks, Adam D. and Hlozek, Ren\'ee and Hubmayr, Johannes and Hughes, John P. and Irwin, Kent D. and Koopman, Brian and Kosowsky, Arthur and Li, Dale and Louis, Thibaut and Lungu, Marius and Madhavacheril, Mathew and Maurin, Lo\"{\i}c and McMahon, Jeffrey John and Moodley, Kavilan and Naess, Sigurd and Nati, Federico and Newburgh, Laura and Niemack, Michael D. and Page, Lyman A. and Pappas, Christine G. and Partridge, Bruce and Schmitt, Benjamin L. and Sehgal, Neelima and Sherwin, Blake D. and Sievers, Jonathan L. and Spergel, David N. and Staggs, Suzanne T. and van Engelen, Alexander and Wollack, Edward J.},
  collaboration = {ACTPol Collaboration},
  journal = {Phys. Rev. D},
  volume = {93},
  issue = {8},
  pages = {082002},
  numpages = {8},
  year = {2016},
  month = {Apr},
  publisher = {American Physical Society},
  doi = {10.1103/PhysRevD.93.082002},
  url = {https://link.aps.org/doi/10.1103/PhysRevD.93.082002}
}

@article{PhysRevD.108.023516,
  title = {Kinematic Sunyaev-Zel'dovich effect with ACT, DES, and BOSS: A novel hybrid estimator},
  author = {Mallaby-Kay, M. and Amodeo, S. and Hill, J. C. and Aguena, M. and Allam, S. and Alves, O. and Annis, J. and Battaglia, N. and Battistelli, E. S. and Baxter, E. J. and Bechtol, K. and Becker, M. R. and Bertin, E. and Bond, J. R. and Brooks, D. and Calabrese, E. and Carnero Rosell, A. and Carrasco Kind, M. and Carretero, J. and Choi, A. and Crocce, M. and da Costa, L. N. and Pereira, M. E. S. and De Vicente, J. and Desai, S. and Dietrich, J. P. and Doel, P. and Doux, C. and Drlica-Wagner, A. and Dunkley, J. and Elvin-Poole, J. and Everett, S. and Ferraro, S. and Ferrero, I. and Frieman, J. and Gallardo, P. A. and Garc\'{\i}a-Bellido, J. and Giannini, G. and Gruen, D. and Gruendl, R. A. and Gutierrez, G. and Hinton, S. R. and Hollowood, D. L. and James, D. J. and Kosowsky, A. and Kuehn, K. and Lokken, M. and Louis, T. and Marshall, J. L. and McMahon, J. and Mena-Fern\'andez, J. and Menanteau, F. and Miquel, R. and Moodley, K. and Mroczkowski, T. and Naess, S. and Niemack, M. D. and Ogando, R. L. C. and Page, L. and Pandey, S. and Pieres, A. and Plazas Malag\'on, A. A. and Raveri, M. and Rodriguez-Monroy, M. and Rykoff, E. S. and Samuroff, S. and Sanchez, E. and Schaan, E. and Sevilla-Noarbe, I. and Sheldon, E. and Sif\'on, C. and Smith, M. and Soares-Santos, M. and Sobreira, F. and Suchyta, E. and Tarle, G. and To, C. and Vargas, C. and Vavagiakis, E. M. and Weaverdyck, N. and Weller, J. and Wiseman, P. and Yanny, B.},
  journal = {Phys. Rev. D},
  volume = {108},
  issue = {2},
  pages = {023516},
  numpages = {19},
  year = {2023},
  month = {Jul},
  publisher = {American Physical Society},
  doi = {10.1103/PhysRevD.108.023516},
  url = {https://link.aps.org/doi/10.1103/PhysRevD.108.023516}
}

@misc{Sunseri:2025hhj,
    author = "Sunseri, James and Amon, Alexandra and Dunkley, Jo and Battaglia, Nicholas and Ferraro, Simone and Hadzhiyska, Boryana and Ried Guachalla, Bernadita and Schaan, Emmanuel",
    title = "{Disentangling the Halo: Joint Model for Measurements of the Kinetic Sunyaev-Zeldovich Effect and Galaxy-Galaxy Lensing}",
    eprint = "2505.20413",
    archivePrefix = "arXiv",
    primaryClass = "astro-ph.CO",
    month = "5",
    year = "2025"
}

@misc{Bigwood:2025kur,
    author = "Bigwood, Leah and Yamamoto, Masaya and Siegel, Jared and Amon, Alexandra and McCarthy, Ian G. and Dave, Romeel and Salcido, Jaime and Schaller, Matthieu and Schaye, Joop and Yang, Tianyi",
    title = "{The kinetic Sunyaev Zeldovich effect as a benchmark for AGN feedback models in hydrodynamical simulations: insights from DESI + ACT}",
    eprint = "2510.15822",
    archivePrefix = "arXiv",
    primaryClass = "astro-ph.CO",
    month = "10",
    year = "2025"
}

@misc{Siegel:2025ivd,
    author = "Siegel, Jared and Bigwood, Leah and Amon, Alexandra and McCullough, Jamie and Yamamoto, Masaya and McCarthy, Ian G. and Schaller, Matthieu and Schneider, Aurel and Schaye, Joop",
    title = "{The suppression of the matter power spectrum: strong feedback from X-ray gas mass fractions, kSZ effect profiles, and galaxy-galaxy lensing}",
    eprint = "2512.02954",
    archivePrefix = "arXiv",
    primaryClass = "astro-ph.CO",
    month = "12",
    year = "2025"
}

@article{DES:2024iny,
    author = "Bigwood, L. and others",
    collaboration = "DES",
    title = "{Weak lensing combined with the kinetic Sunyaev{\textendash}Zel{\textquoteright}dovich effect: a study of baryonic feedback}",
    eprint = "2404.06098",
    archivePrefix = "arXiv",
    primaryClass = "astro-ph.CO",
    reportNumber = "DES-2024-0827, FERMILAB-PUB-24-0130-PPD",
    doi = "10.1093/mnras/stae2100",
    journal = "Mon. Not. Roy. Astron. Soc.",
    volume = "534",
    number = "1",
    pages = "655--682",
    year = "2024"
}

@article{PhysRevLett.117.051301,
  title = {Kinematic Sunyaev-Zel'dovich Effect with Projected Fields: A Novel Probe of the Baryon Distribution with Planck, WMAP, and WISE Data},
  author = {Hill, J. Colin and Ferraro, Simone and Battaglia, Nick and Liu, Jia and Spergel, David N.},
  journal = {Phys. Rev. Lett.},
  volume = {117},
  issue = {5},
  pages = {051301},
  numpages = {6},
  year = {2016},
  month = {Jul},
  publisher = {American Physical Society},
  doi = {10.1103/PhysRevLett.117.051301},
  url = {https://link.aps.org/doi/10.1103/PhysRevLett.117.051301}
}

@article{PhysRevD.94.123526,
  title = {Kinematic Sunyaev-Zel'dovich effect with projected fields. II. Prospects, challenges, and comparison with simulations},
  author = {Ferraro, Simone and Hill, J. Colin and Battaglia, Nick and Liu, Jia and Spergel, David N.},
  journal = {Phys. Rev. D},
  volume = {94},
  issue = {12},
  pages = {123526},
  numpages = {14},
  year = {2016},
  month = {Dec},
  publisher = {American Physical Society},
  doi = {10.1103/PhysRevD.94.123526},
  url = {https://link.aps.org/doi/10.1103/PhysRevD.94.123526}
}

@article{PhysRevD.104.043518,
  title = {Constraining the baryon abundance with the kinematic Sunyaev-Zel'dovich effect: Projected-field detection using $Planck$, $WMAP$, and $unWISE$},
  author = {Kusiak, Aleksandra and Bolliet, Boris and Ferraro, Simone and Hill, J. Colin and Krolewski, Alex},
  journal = {Phys. Rev. D},
  volume = {104},
  issue = {4},
  pages = {043518},
  numpages = {28},
  year = {2021},
  month = {Aug},
  publisher = {American Physical Society},
  doi = {10.1103/PhysRevD.104.043518},
  url = {https://link.aps.org/doi/10.1103/PhysRevD.104.043518}
}

@misc{Lague:2025txe,
    author = {Lagu{\"e}, Alex and Madhavacheril, Mathew S. and Borrow, Josh and Smith, Kendrick M. and Chen, Xinyi and Schaller, Matthieu and Schaye, Joop},
    title = "{Inferring the Impacts of Baryonic Feedback from Kinetic Sunyaev-Zeldovich Cross-Correlations}",
    eprint = "2511.20595",
    archivePrefix = "arXiv",
    primaryClass = "astro-ph.CO",
    month = "11",
    year = "2025"
}

@misc{mccarthy2025flamingocombiningkineticsz,
      title={FLAMINGO: combining kinetic SZ effect and galaxy-galaxy lensing measurements to gauge the impact of feedback on large-scale structure}, 
      author={Ian G. McCarthy and Alexandra Amon and Joop Schaye and Emmanuel Schaan and Raul E. Angulo and Jaime Salcido and Matthieu Schaller and Leah Bigwood and Willem Elbers and Roi Kugel and John C. Helly and Victor J. Forouhar Moreno and Carlos S. Frenk and Robert J. McGibbon and Lurdes Ondaro-Mallea and Marcel P. van Daalen},
      year={2025},
      eprint={2410.19905},
      archivePrefix={arXiv},
      primaryClass={astro-ph.CO},
      url={https://arxiv.org/abs/2410.19905}, 
}

@misc{salcido2025implicationsfeedbacksolutionss8,
      title={Implications of feedback solutions to the $S_8$ tension for the baryon fractions of galaxy groups and clusters}, 
      author={Jaime Salcido and Ian G. McCarthy},
      year={2025},
      eprint={2409.05716},
      archivePrefix={arXiv},
      primaryClass={astro-ph.CO},
      url={https://arxiv.org/abs/2409.05716}, 
}

@misc{harscouet2025kszeveryonepseudoclapproach,
      title={kSZ for everyone: the pseudo-Cl approach to stacking}, 
      author={Lea Harscouet and Kevin Wolz and Amy Wayland and David Alonso and Boryana Hadzhiyska},
      year={2025},
      eprint={2512.14625},
      archivePrefix={arXiv},
      primaryClass={astro-ph.CO},
      url={https://arxiv.org/abs/2512.14625}, 
}

@misc{hotinli2025velocityreconstructionkszmeasuring,
      title={Velocity Reconstruction from KSZ: Measuring $f_{NL}$ with ACT and DESILS}, 
      author={Selim C. Hotinli and Kendrick M. Smith and Simone Ferraro},
      year={2025},
      eprint={2506.21657},
      archivePrefix={arXiv},
      primaryClass={astro-ph.CO},
      url={https://arxiv.org/abs/2506.21657}, 
}

@misc{lai2025kszvelocityreconstructionact,
      title={KSZ Velocity Reconstruction with ACT and DESI-LS using a Tomographic QML Power Spectrum Estimator}, 
      author={Anderson C. M. Lai and Yurii Kvasiuk and Moritz Münchmeyer},
      year={2025},
      eprint={2506.21684},
      archivePrefix={arXiv},
      primaryClass={astro-ph.CO},
      url={https://arxiv.org/abs/2506.21684}, 
}

@misc{lague2024constraintslocalprimordialnongaussianity,
      title={Constraints on local primordial non-Gaussianity with 3d Velocity Reconstruction from the Kinetic Sunyaev-Zeldovich Effect}, 
      author={Alex Laguë and Mathew S. Madhavacheril and Kendrick M. Smith and Simone Ferraro and Emmanuel Schaan},
      year={2024},
      eprint={2411.08240},
      archivePrefix={arXiv},
      primaryClass={astro-ph.CO},
      url={https://arxiv.org/abs/2411.08240}, 
}

@misc{mccarthy2025atacamacosmologytelescopecrosscorrelation,
      title={The Atacama Cosmology Telescope: Cross-correlation of kSZ and continuity equation velocity reconstruction with photometric DESI LRGs}, 
      author={Fiona McCarthy and Boryana Hadzhiyska and J. Richard Bond and William R. Coulton and Jo Dunkley and Carmen Embil Villagra and Matthew C. Johnson and Kavilan Moodley and Toshiya Namikawa and Bernardita Ried Guachalla and Blake D. Sherwin and Cristóbal Sifón and Alexander van Engelen and Eve M. Vavagiakis and Edward J. Wollack},
      year={2025},
      eprint={2511.15701},
      archivePrefix={arXiv},
      primaryClass={astro-ph.CO},
      url={https://arxiv.org/abs/2511.15701}, 
}

@misc{mccarthy2024atacamacosmologytelescopelargescale,
      title={The Atacama Cosmology Telescope: Large-scale velocity reconstruction with the kinematic Sunyaev--Zel'dovich effect and DESI LRGs}, 
      author={Fiona McCarthy and Nicholas Battaglia and Rachel Bean and J. Richard Bond and Hongbo Cai and Erminia Calabrese and William R. Coulton and Mark J. Devlin and Jo Dunkley and Simone Ferraro and Vera Gluscevic and Yilun Guan and J. Colin Hill and Matthew C. Johnson and Aleksandra Kusiak and Alex Laguë and Niall MacCrann and Mathew S. Madhavacheril and Kavilan Moodley and Sigurd Naess and Frank J. Qu and Bernardita Ried Guachalla and Neelima Sehgal and Blake D. Sherwin and Cristóbal Sifón and Kendrick M. Smith and Suzanne T. Staggs and Alexander van Engelen and Eve M. Vavagiakis and Edward J. Wollack},
      year={2024},
      eprint={2410.06229},
      archivePrefix={arXiv},
      primaryClass={astro-ph.CO},
      url={https://arxiv.org/abs/2410.06229}, 
}

@article{Schneider_2022,
   title={Constraining baryonic feedback and cosmology with weak-lensing, X-ray, and kinematic Sunyaev–Zeldovich observations},
   volume={514},
   ISSN={1365-2966},
   url={http://dx.doi.org/10.1093/mnras/stac1493},
   DOI={10.1093/mnras/stac1493},
   number={3},
   journal={Monthly Notices of the Royal Astronomical Society},
   publisher={Oxford University Press (OUP)},
   author={Schneider, Aurel and Giri, Sambit K and Amodeo, Stefania and Refregier, Alexandre},
   year={2022},
   month=jun, pages={3802–3814} }

@misc{bigwood2025kineticsunyaevzeldovicheffect,
      title={The kinetic Sunyaev Zeldovich effect as a benchmark for AGN feedback models in hydrodynamical simulations: insights from DESI + ACT}, 
      author={Leah Bigwood and Masaya Yamamoto and Jared Siegel and Alexandra Amon and Ian G. McCarthy and Romeel Dave and Jaime Salcido and Matthieu Schaller and Joop Schaye and Tianyi Yang},
      year={2025},
      eprint={2510.15822},
      archivePrefix={arXiv},
      primaryClass={astro-ph.CO},
      url={https://arxiv.org/abs/2510.15822}, 
}

@misc{kovac2025baryonificationiiconstrainingfeedback,
      title={Baryonification II: Constraining feedback with X-ray and kinematic Sunyaev-Zel'dovich observations}, 
      author={Michael Kovač and Andrina Nicola and Jozef Bucko and Aurel Schneider and Robert Reischke and Sambit K. Giri and Romain Teyssier and Matthieu Schaller and Joop Schaye},
      year={2025},
      eprint={2507.07991},
      archivePrefix={arXiv},
      primaryClass={astro-ph.CO},
      url={https://arxiv.org/abs/2507.07991}, 
}

@ARTICLE{2026arXiv260419744Q,
       author = {{Qu}, F.~J. and {Ried Guachalla}, B. and {Schaan}, E. and {Hadzhiyska}, B. and {Ferraro}, S. and {Aguilar}, J. and {Ahlen}, S. and {Baleato Lizancos}, A. and {Bianchi}, D. and {Brooks}, D. and {Canning}, R. and {Castander}, F.~J. and {Chaussidon}, E. and {Claybaugh}, T. and {Cuceu}, A. and {de la Macorra}, A. and {Dey}, B. and {Doel}, P. and {Font-Ribera}, A. and {Forero-Romero}, J.~E. and {Gazta{\~n}aga}, E. and {Gontcho}, S. Gontcho A and {Gutierrez}, G. and {Herrera-Alcantar}, H.~K. and {Honscheid}, K. and {Howlett}, C. and {Huterer}, D. and {Ishak}, M. and {Kehoe}, R. and {Kisner}, T. and {Kremin}, A. and {Lahav}, O. and {Landriau}, M. and {Le Guillou}, L. and {Levi}, M.~E. and {Manera}, M. and {Meisner}, A. and {Miquel}, R. and {Nadathur}, S. and {Newman}, J.~A. and {Percival}, W.~J. and {P'erez-R`afols}, I. and {Rossi}, G. and {Samushia}, L. and {Sanchez}, E. and {Schlafly}, E.~F. and {Schlegel}, D. and {Schubnell}, M. and {Seo}, H. and {Silber}, J. and {Sprayberry}, D. and {Tarl'e}, G. and {Weaver}, B.~A. and {Zhou}, R.},
        title = "{Precision Kinematic Sunyaev--Zel'dovich Measurements Across Halo Mass and Redshift with DESI DR2 and ACT DR6: Part I. Luminous Red Galaxies}",
      journal = {arXiv e-prints},
         year = 2026,
        month = apr,
          eid = {arXiv:2604.19744},
        pages = {arXiv:2604.19744},
          doi = {10.48550/arXiv.2604.19744},
archivePrefix = {arXiv},
       eprint = {2604.19744},
 primaryClass = {astro-ph.CO},
       adsurl = {https://ui.adsabs.harvard.edu/abs/2026arXiv260419744Q}
}

@ARTICLE{2026arXiv260419745H,
       author = {{Hadzhiyska}, B. and {Ferraro}, S. and {Qu}, F.~J. and {Ried Guachalla}, B. and {Schaan}, E. and {Aguilar}, J. and {Ahlen}, S. and {Bianchi}, D. and {Brooks}, D. and {Castander}, F.~J. and {Chaussidon}, E. and {Claybaugh}, T. and {de la Macorra}, A. and {Dey}, Arjun and {Dey}, Biprateep and {Doel}, P. and {Forero-Romero}, J.~E. and {Gazta{\~n}aga}, E. and {Gontcho}, S. Gontcho A and {Gutierrez}, G. and {Guy}, J. and {Honscheid}, K. and {Howlett}, C. and {Huterer}, D. and {Ishak}, M. and {Joyce}, R. and {Kehoe}, R. and {Kisner}, T. and {Kremin}, A. and {Lahav}, O. and {Landriau}, M. and {Le Guillou}, L. and {Leauthaud}, A. and {Manera}, M. and {Martini}, P. and {Meisner}, A. and {Miquel}, R. and {Nadathur}, S. and {Palanque-Delabrouille}, N. and {Percival}, W.~J. and {Prada}, F. and {P{\'e}rez-R{\`a}fols}, I. and {Rossi}, G. and {Samushia}, L. and {Sanchez}, E. and {Schlafly}, E.~F. and {Schlegel}, D. and {Silber}, J. and {Sprayberry}, D. and {Tarl{\'e}}, G. and {Weaver}, B.~A. and {Zhou}, R. and {Zou}, H.},
        title = "{Precision Kinematic Sunyaev--Zel'dovich Measurements Across Halo Mass and Redshift with DESI DR2 and ACT DR6: Part II. Bright Galaxy Survey and Emission-Line Galaxies}",
      journal = {arXiv e-prints},
         year = 2026,
        month = apr,
          eid = {arXiv:2604.19745},
        pages = {arXiv:2604.19745},
          doi = {10.48550/arXiv.2604.19745},
archivePrefix = {arXiv},
       eprint = {2604.19745},
 primaryClass = {astro-ph.CO},
       adsurl = {https://ui.adsabs.harvard.edu/abs/2026arXiv260419745H}
}

@ARTICLE{2025arXiv250919514L,
       author = {{Leung}, Calvin and {Borrow}, Josh and {Masui}, Kiyoshi W. and {Andrew}, Shion and {Chen}, Kai-Feng and {Schaye}, Joop and {Schaller}, Matthieu},
        title = "{Nulling baryonic feedback in weak lensing surveys using cross-correlations with fast radio bursts}",
      journal = {arXiv e-prints},
         year = 2025,
        month = sep,
          eid = {arXiv:2509.19514},
        pages = {arXiv:2509.19514},
          doi = {10.48550/arXiv.2509.19514},
archivePrefix = {arXiv},
       eprint = {2509.19514},
 primaryClass = {astro-ph.CO},
       adsurl = {https://ui.adsabs.harvard.edu/abs/2025arXiv250919514L}
}

@ARTICLE{2026arXiv260304397K,
       author = {{Kadir}, Sadaf and {Ried Guachalla}, Bernardita and {Yuan}, Sihan and {Schaan}, Emmanuel and {Wechsler}, Risa H.},
        title = "{Exploring gas thermodynamics around galaxies from the Sunyaev-Zel'dovich effects: impact of galaxy-halo connection, 2D projection and velocity field}",
      journal = {arXiv e-prints},
         year = 2026,
        month = mar,
          eid = {arXiv:2603.04397},
        pages = {arXiv:2603.04397},
          doi = {10.48550/arXiv.2603.04397},
archivePrefix = {arXiv},
       eprint = {2603.04397},
 primaryClass = {astro-ph.CO},
       adsurl = {https://ui.adsabs.harvard.edu/abs/2026arXiv260304397K}
}

@ARTICLE{2021ApJ...919....2M,
       author = {{Moser}, Emily and {Amodeo}, Stefania and {Battaglia}, Nicholas and {Alvarez}, Marcelo A. and {Ferraro}, Simone and {Schaan}, Emmanuel},
        title = "{The Impacts of Modeling Choices on the Inference of Circumgalactic Medium Properties from Sunyaev-Zeldovich Observations}",
      journal = {\apj},
         year = 2021,
        month = sep,
       volume = {919},
       number = {1},
          eid = {2},
        pages = {2},
          doi = {10.3847/1538-4357/ac0cea},
archivePrefix = {arXiv},
       eprint = {2103.02469},
 primaryClass = {astro-ph.GA},
       adsurl = {https://ui.adsabs.harvard.edu/abs/2021ApJ...919....2M}
}

@ARTICLE{2023arXiv230710919M,
       author = {{Moser}, Emily and {Battaglia}, Nicholas and {Amodeo}, Stefania},
        title = "{Searching for Systematics in Forward Modeling Sunyaev-Zeldovich Profiles}",
      journal = {arXiv e-prints},
         year = 2023,
        month = jul,
          eid = {arXiv:2307.10919},
        pages = {arXiv:2307.10919},
          doi = {10.48550/arXiv.2307.10919},
archivePrefix = {arXiv},
       eprint = {2307.10919},
 primaryClass = {astro-ph.CO},
       adsurl = {https://ui.adsabs.harvard.edu/abs/2023arXiv230710919M}
}

@article{SO:ScienceGoalsForecasts,
  author        = {Ade, P. and others},
  collaboration = {Simons Observatory},
  title         = {The Simons Observatory: Science goals and forecasts},
  eprint        = {1808.07445},
  archivePrefix = {arXiv},
  primaryClass  = {astro-ph.CO},
  journal       = {JCAP},
  volume        = {2019},
  number        = {02},
  pages         = {056},
  year          = {2019},
  doi           = {10.1088/1475-7516/2019/02/056}
}

@misc{CMBHD:SnowmassWP,
  author        = {Sehgal, Neelima and others},
  title         = {Snowmass2021 CMB-HD White Paper},
  eprint        = {2203.05728},
  archivePrefix = {arXiv},
  primaryClass  = {astro-ph.CO},
  year          = {2022},
  doi           = {10.48550/arXiv.2203.05728}
}

@article{Hadzhiyska2026,
  author  = {Hadzhiyska, Boryana},
  title   = {Shear-kSZ: A New Estimator for the Matter-Electron Power
             Spectrum from kSZ Tomography and Weak Lensing},
  journal = {arXiv e-prints},
  year    = {2026},
  eprint  = {2607.27149},
  archivePrefix = {arXiv},
  primaryClass  = {astro-ph.CO}
}

@ARTICLE{Ivezic2019,
  author = {{Ivezi{\'c}}, {\v{Z}}. and {Kahn}, S.~M. and {Tyson}, J.~A. and others},
  title = "{LSST: From Science Drivers to Reference Design and Anticipated Data Products}",
  journal = {ApJ},
  year = 2019,
  volume = {873},
  number = {2},
  pages = {111},
  doi = {10.3847/1538-4357/ab042c},
  eprint = {0805.2366}
}

@ARTICLE{Laureijs2011,
  author = {{Laureijs}, R. and {Amiaux}, J. and {Arduini}, S. and others},
  title = "{Euclid Definition Study Report}",
  journal = {arXiv e-prints},
  year = 2011,
  eprint = {1110.3193}
}

@ARTICLE{Spergel2015,
  author = {{Spergel}, D. and {Gehrels}, N. and {Baltay}, C. and others},
  title = "{Wide-Field InfrarRed Survey Telescope-Astrophysics Focused Telescope Assets WFIRST-AFTA 2015 Report}",
  journal = {arXiv e-prints},
  year = 2015,
  eprint = {1503.03757}
}

@article{Schaye2023,
  author  = {Schaye, Joop and Kugel, Roi and Schaller, Matthieu and
             Helly, John C. and Braspenning, Joey and Elbers, Willem and
             McCarthy, Ian G. and van Daalen, Marcel P. and Vandenbroucke, Bert
             and Frenk, Carlos S. and others},
  title   = {The {FLAMINGO} project: cosmological hydrodynamical simulations
             for large-scale structure and galaxy cluster surveys},
  journal = {MNRAS},
  volume  = {526},
  number  = {4},
  pages   = {4978--5020},
  year    = {2023},
  doi     = {10.1093/mnras/stad2419},
  eprint  = {2306.04024},
  archivePrefix = {arXiv},
  primaryClass  = {astro-ph.CO}
}

@article{Mead2020,
  author  = {Mead, A.~J. and Brieden, S. and Tr\"oster, T. and Heymans, C.},
  title   = {{HMcode-2020}: improved modelling of non-linear cosmological power
             spectra with baryonic feedback},
  journal = {MNRAS},
  volume  = {502},
  number  = {1},
  pages   = {1401--1422},
  year    = {2021},
  doi     = {10.1093/mnras/stab082},
  eprint  = {2009.01858},
  archivePrefix = {arXiv},
  primaryClass  = {astro-ph.CO}
}

@article{vanDaalen2020,
  author  = {van Daalen, Marcel P. and McCarthy, Ian G. and Schaye, Joop},
  title   = {Exploring the effects of galaxy formation on matter clustering
             through a library of simulation power spectra},
  journal = {MNRAS},
  volume  = {491},
  number  = {2},
  pages   = {2424--2446},
  year    = {2020},
  doi     = {10.1093/mnras/stz3199},
  eprint  = {1906.00968},
  archivePrefix = {arXiv},
  primaryClass  = {astro-ph.CO}
}

@article{SchneiderTeyssier2015,
  author  = {Schneider, Aurel and Teyssier, Romain},
  title   = {A new method to quantify the effects of baryons on the matter
             power spectrum},
  journal = {JCAP},
  volume  = {2015},
  number  = {12},
  pages   = {049},
  year    = {2015},
  doi     = {10.1088/1475-7516/2015/12/049},
  eprint  = {1510.06034},
  archivePrefix = {arXiv},
  primaryClass  = {astro-ph.CO}
}

@article{Arico2021,
  author  = {Aric\`o, Giovanni and Angulo, Raul E. and Contreras, Sergio and
             Ondaro-Mallea, Lurdes and Pellejero-Iba\~nez, Marcos and
             Zennaro, Matteo},
  title   = {The {BACCO} simulation project: a baryonification emulator with
             neural networks},
  journal = {MNRAS},
  volume  = {506},
  number  = {3},
  pages   = {4070--4082},
  year    = {2021},
  doi     = {10.1093/mnras/stab1911},
  eprint  = {2011.15018},
  archivePrefix = {arXiv},
  primaryClass  = {astro-ph.CO}
}

@ARTICLE{Vogelsberger2020,
       author = {{Vogelsberger}, Mark and {Marinacci}, Federico and {Torrey}, Paul and {Puchwein}, Ewald},
        title = "{Cosmological simulations of galaxy formation}",
      journal = {Nature Reviews Physics},
         year = 2020,
        month = jan,
       volume = {2},
       number = {1},
        pages = {42-66},
          doi = {10.1038/s42254-019-0127-2},
archivePrefix = {arXiv},
       eprint = {1909.07976},
 primaryClass = {astro-ph.GA},
       adsurl = {https://ui.adsabs.harvard.edu/abs/2020NatRP...2...42V}
}

@ARTICLE{Crain2023,
       author = {{Crain}, Robert A. and {van de Voort}, Freeke},
        title = "{Hydrodynamical Simulations of the Galaxy Population: Enduring Successes and Outstanding Challenges}",
      journal = {\araa},
         year = 2023,
        month = aug,
       volume = {61},
        pages = {473-515},
          doi = {10.1146/annurev-astro-041923-043618},
archivePrefix = {arXiv},
       eprint = {2309.17075},
 primaryClass = {astro-ph.GA},
       adsurl = {https://ui.adsabs.harvard.edu/abs/2023ARA&A..61..473C}
}

@ARTICLE{Navarro2021,
       author = {{Villaescusa-Navarro}, Francisco and {Angl{\'e}s-Alc{\'a}zar}, Daniel and {Genel}, Shy and {Spergel}, David N. and {Somerville}, Rachel S. and {Dave}, Romeel and {Pillepich}, Annalisa and {Hernquist}, Lars and {Nelson}, Dylan and {Torrey}, Paul and {Narayanan}, Desika and {Li}, Yin and {Philcox}, Oliver and {La Torre}, Valentina and {Maria Delgado}, Ana and {Ho}, Shirley and {Hassan}, Sultan and {Burkhart}, Blakesley and {Wadekar}, Digvijay and {Battaglia}, Nicholas and {Contardo}, Gabriella and {Bryan}, Greg L.},
        title = "{The CAMELS Project: Cosmology and Astrophysics with Machine-learning Simulations}",
      journal = {\apj},
         year = 2021,
        month = jul,
       volume = {915},
       number = {1},
          eid = {71},
        pages = {71},
          doi = {10.3847/1538-4357/abf7ba},
archivePrefix = {arXiv},
       eprint = {2010.00619},
 primaryClass = {astro-ph.CO},
       adsurl = {https://ui.adsabs.harvard.edu/abs/2021ApJ...915...71V}
}
\bibliographystyle{unsrtads}
\end{document}